 \documentclass[final,3p,times]{elsarticle}

\usepackage{amssymb}
\usepackage{babel,tabularx,ragged2e,booktabs}

\usepackage{lineno}

\usepackage[utf8]{inputenc}
\usepackage{textcomp}
\usepackage[mathscr]{euscript}
\usepackage{graphicx}
\usepackage{amsmath}
\usepackage[version=4]{mhchem}
\usepackage{siunitx}
\usepackage{longtable,tabularx}
\usepackage{color}
\usepackage{cancel}
\usepackage{gensymb}
\usepackage{color}
\usepackage{tcolorbox}
\usepackage{float}
\usepackage{multirow}
\usepackage{caption}
\usepackage{subcaption}
\usepackage{graphicx}
\usepackage{amsmath}
\usepackage[colorlinks]{hyperref}

\colorlet{colorRev1}{blue!80!black} 
\colorlet{colorRev2}{red!80!black} 

\journal{Aerospace Science and Technology}

\begin{document}

\begin{frontmatter}



\title{Modeling, robust control synthesis and worst-case analysis for an on-orbit servicing mission with large flexible spacecraft}


\author[ISAE]{R. Rodrigues\corref{cor1}}
\ead{ricardo.rodrigues@isae-supaero.fr}
\cortext[cor1]{Corresponding author; Ph.D. student.}

\author[ESA]{V. Preda}
\ead{valentin.preda@esa.int}

\author[ISAE]{F. Sanfedino\fnref{label2}}
\ead{francesco.sanfedino@isae-supaero.fr}
\fntext[label2]{Associate Professor}

\author[ISAE]{D. Alazard\fnref{label3}}
\ead{daniel.alazard@isae-supaero.fr}
\fntext[label3]{Professor}

\address[ISAE]{Institut Supérieur de l’Aéronautique et de l’Espace (ISAE-SUPAERO), Université de Toulouse, 10 Avenue Edouard Belin, BP-54032, 31055, Toulouse
Cedex 4, France}
\address[ESA]{ESA/ESTEC, Keplerlaan 1, 2201 AZ, Noordwijk, The Netherlands}

\begin{abstract}

This paper outlines a complete methodology for modeling an on-orbit servicing mission scenario and designing a feedback control system for the attitude dynamics that is guaranteed to robustly meet pointing requirements, despite model uncertainties as well as large inertia and flexibility changes throughout the mission scenario. A model of the uncertain plant was derived, which fully captures the dynamics and couplings between all subsystems as well as the decoupled/coupled configurations of the chaser/target system in a single linear fractional representation (LFR). In addition, a new approach is proposed to model and analyze a closed-loop kinematic chain formed by the chaser and the target spacecraft through the chaser's robotic arm, which uses two local spring-damper systems with uncertain damping and stiffness. This approach offers the possibility to model the dynamical behaviour of a docking mechanism with dynamic stiffness and damping. The controller was designed by taking into account all the interactions between subsystems and uncertainties as well as the time-varying and coupled flexible dynamics. Lastly, the robust stability and worst-case performances were assessed by means of a structured singular value analysis.
\end{abstract}


\end{frontmatter}

\nolinenumbers


\section{Introduction}

\subsection{Background and motivation}

On the one hand, the autonomous assembly of large structures in space is a key challenge for missions possessing structures to be self-deployed as a single piece, like PULSAR, which is able to overcome the size restrictions of current launchers \cite{Cumer2021, Deremetz2020, Rognant2019}. On the other hand, there are also many ongoing on-orbit servicing (OOS) projects, which fall into three main categories: observation, motion and manipulation \cite{Cr2018}. Both ESA’s Clean Space initiative and NASA’s Orbital Debris Program Office are mainly focused on de-orbiting satellites, while investigating the capture of a satellite by means of space robotics \cite{Wormnes2013}. However, the use of robotics can also be a solution for different types of OOS missions such as maintenance, repair, refuel, upgrade and docked inspection of a satellite, which come under the manipulation category. NGIS' MEV-1 \cite{MEP2} and EROSS \cite{Dubanchet2020, Dubanchet2020a, Andiappane2019}  are some of the examples. 

Over the years, a wide variety of studies has been conducted on the topic of spacecraft on-orbit rendezvous and assembly with disturbance rejection \cite{LIU2022107358,Biggs2021,Colagrossi2021,Samsam2022,CHAI2020105894,Henry2021}. However, most of these studies neglect flexibility and study the system as a rigid mass. Moreover, some papers consider the presence of uncertainties when doing control design \cite{Jonchay2021}. When it comes to study the docking with another satellite on-orbit, some approaches acknowledge the capture mechanism as a simple basket \cite{Henry2021}, while others consider the usage of a robotic arm for this purpose \cite{papadoulos2007, Brannan2013, RAINA202121, Wang2006}. Nevertheless, these studies do not analyze in depth the change in mechanical properties due to the robotic arm movement while being docked to a target spacecraft with flexible appendages and considerable mass/inertial properties. Several docking mechanisms such as HOTDOCK \cite{Deremetz2020} or ASSIST \cite{Medina2017} are also being developed in order to make OOS missions possible. The contact dynamics model of ASSIST is built in a simulator \cite{Medina2017} by considering the forces and torques caused by physical contact between the chaser and target spacecraft. However, no symbolic linear model is obtained in order to consider this effect when designing a controller. 

From an AOCS/GNC point of view, this type of complex missions is particularly challenging due to the time-varying \& coupled flexible dynamics. Consequently, the success of these projects is constrained by the ability to have an accurate model of the system and analysis tools which allow to predict the worst-case scenarios during preliminary design phases.

In this context, this paper aims at proposing an end-to-end structure/control co-design activity for an on-orbit servicing mission scenario. The ambition is to fill an important gap in the literature by taking into account flexibility and system uncertainty in the design of a robust controller to be used in an orbital servicing operation. In fact, this kind of scenario will be more and more frequent in the near future of space missions, where on-orbit operations of large and flexible structures will be put in place. In this context, a full modeling of multibody complex structures can help predicting in a very early phase the worst-case scenarios and push the control system to its limits of performance. Moreover, the availability of a model valid for any possible configuration simplifies the synthesis of the controller, which does not have to switch from one control mode to another. The transition between different control phases is in fact a critical aspect of the control design, that often implies intermediate tranquilization time windows.

The goal is then to establish a methodology for performing robust control system design and worst-case analysis for an on-orbit servicing scenario involving the interaction between two flexible vehicles. A Linear Parameter-Varying (LPV) model of two flexible multibody systems was firstly obtained. This model is fully parameterized according to the system's geometrical configuration and was built using the Two-Input Two-Output Ports (TITOP) approach \cite{Sanfedino2018,Chebbi2017,sanfe2019,SANFEDINO2022108168,Gonzalez2016,Perez2016}, which offers the possibility to model complex multibody mechanical systems, while keeping the uncertain nature of the plant and condensing all the possible mechanical configurations in a single low order LFR. This model is then ready for robust control synthesis as well as robust stability and performance assessment \cite{Preda2020}. All the models derived with the TITOP approach have been systematically implemented in the last release of the Satellite Dynamics Toolbox (SDT) \cite{userguide}, which allows the user to easily build the model of a flexible spacecraft with several appendages by assembling elemental \textit{Simulink} customized blocks.

The chosen scenario is a system composed of two different spacecraft, a chaser and a target, both with large flexible solar arrays. The target, also called client satellite, is considered collaborative and prepared as it should be designed with specific features to facilitate the rendezvous and capture. First, the chaser performs a rendezvous with the target. Subsequently, the collaborative target is grasped by means of a robotic arm. Finally, the target vehicle is rigidly attached to one of the docking ports on the chaser to free up the robotic manipulator. This type of mission could be used to perform some maintenance on the target spacecraft, but also to use this target as a mission extension pod. It should be noted that this is just a scenario chosen to demonstrate the capabilities of the proposed approach. A multitude of other mission concepts such as on-orbit assembly or refueling could be explored using the same toolset.

\subsection{Contributions and paper organization}

The paper introduces the following key contributions:

\begin{itemize}

\item the development of a complete model fully capturing the dynamics and interactions between all subsystems of an OOS scenario: robotic arm, flexible appendages, decoupled/coupled configurations in a single LFR. This model includes the various interactions and uncertainty effects in a very compact representation. The fact that this model is minimal is critical to allow the usage of modern controller synthesis and analysis tools, which call for reduced numerical complexity.

\item a new approach to the modeling of a closed-loop kinematic chain, which uses an uncertain local damping and stiffness in order to model the dynamical behaviour of a docking mechanism.

\item a thorough controller synthesis and analysis procedure for the design of a static and structured attitude controller, taking into account all the various interactions and couplings which exist between different subsystems.

\end{itemize}

This paper is organized into three parts: system modeling, controller design and stability and performance analysis. In the first part (section \ref{modeling}), a symbolic linear model for control synthesis is obtained using SDT. The result is an LFR, minimal in terms of parameter occurrences. The proposed modeling design is verified using a non-linear physics simulator built with the \textit{Simscape} multibody toolset from \textit{Mathworks}. In the second part of the paper (section \ref{control}), the SDT model is used to design and optimize a controller capable of complying with the performance requirements which are imposed as constraints on the feedback loop. Finally, the third part (section \ref{analysis}) details the rigorous analysis procedure that was used to obtain robust performance and robust stability certificates.

\section{Multibody Modeling Approach}
\label{modeling}

\subsection{The TITOP approach}

\begin{figure}[!ht]
\centering
 \includegraphics[width=1\textwidth]{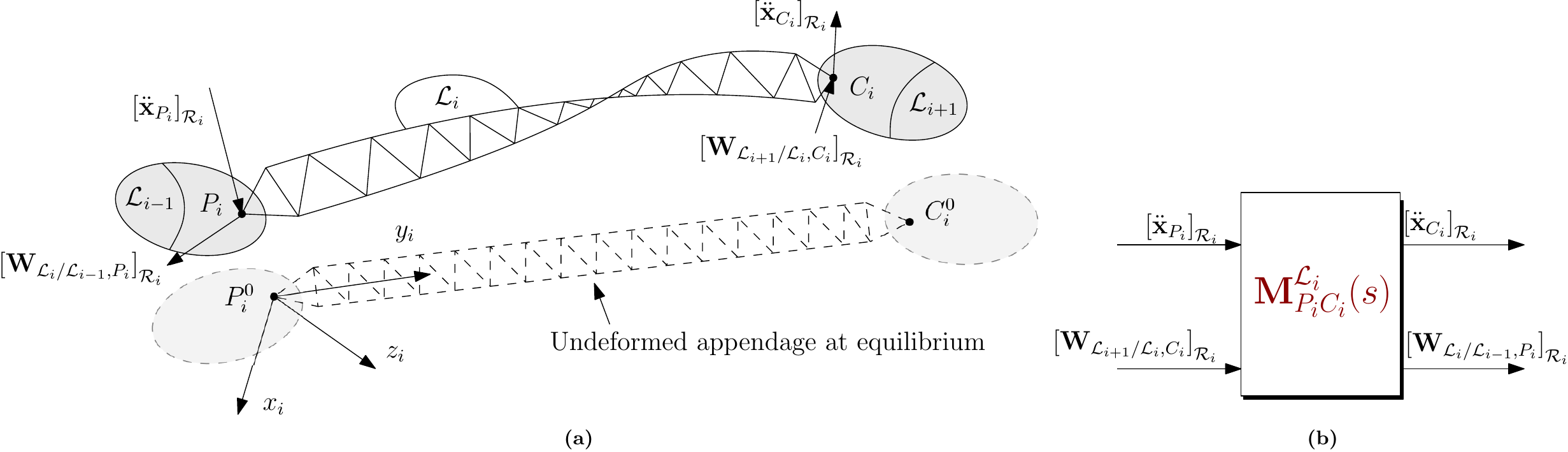}
\caption{(a) i-th flexible appendage of a complex sub-structured body. (b) TITOP model $\mathbf{M}_{P_{i}, C_{i}}^{\mathcal{L}_{i}}(s)$ block-diagram.}
\label{TITOP} 
\end{figure}

The link $\mathcal{L}_{i}$ connected to the parent substructure $\mathcal{L}_{i-1}$ at the point $P_{i}$ and to the child substructure $\mathcal{L}_{i+1}$ at point $C_{i}$ is depicted in Fig. \ref{TITOP}a. The double-port or TITOP model $\mathbf{M}_{P_{i}, C_{i}}^{\mathcal{L}_{i}}(s)$ is a linear dynamic model between 12 inputs:
\begin{itemize}
    
\item the six components in $\mathcal{R}_{i}=\left({P_{i}^0}; x_{i}, y_{i}, z_{i}\right)$ of the wrench ${[\mathbf{W}_{\mathcal{L}_{i+1}/\mathcal{L}_{i},{C_{i}}}]}_{\mathcal{R}_{i}}=\left[\begin{array}{c} \mathbf{F}_{\mathcal{L}_{i+1}/\mathcal{L}_{i},{C_{i}}} \\ \mathbf{T}_{\mathcal{L}_{i+1}/\mathcal{L}_{i},{C_{i}}} \end{array}\right]_{\mathcal{R}_{i}}$ applied by the substructure $\mathcal{L}_{i+1}$ to the link $\mathcal{L}_{i}$ at point $C_{i}: \mathbf{F}_{\mathcal{L}_{i+1}/\mathcal{L}_{i},{C_{i}}}$ stands for the three-component force vector applied at point $C_{i}$, and $\mathbf{T}_{\mathcal{L}_{i+1}/\mathcal{L}_{i},{C_{i}}}$ stands for the three-component torque vector applied at point $C_{i}$.
\item the six components in $\mathcal{R}_{i}$ of the acceleration twist ${[\ddot{\mathbf{x}}_{P_{i}}]}_{\mathcal{R}_{i}}=\left[\begin{array}{c} \mathbf{a}_{P_{i}} \\ \boldsymbol{\dot{\omega}}_{P_{i}}\end{array}\right]_{\mathcal{R}_{i}}$ of point $P_{i}$ : $\mathbf{a}_{P_{i}}$ stands for the three-component linear acceleration vector at point $P_{i}$, and $\boldsymbol{\dot{\omega}}_{P_{i}}$ stands for the three-component angular acceleration vector at point $P_{i}$.
\end{itemize}
and 12 outputs:
\begin{itemize}
\item the six components in $\mathcal{R}_{i}$ of the acceleration twist ${[\ddot{\mathbf{x}}_{C_{i}}]}_{\mathcal{R}_{i}}=\left[\begin{array}{c} \mathbf{a}_{C_{i}} \\ \boldsymbol{\dot{\omega}}_{C_{i}}\end{array}\right]_{\mathcal{R}_{i}}$.
\item the six components in $\mathcal{R}_{i}$ of the wrench ${[\mathbf{W}_{\mathcal{L}_i/\mathcal{L}_{i-1},{P_{i}}}]}_{\mathcal{R}_{i}}=\left[\begin{array}{c} \mathbf{F}_{\mathcal{L}_i/\mathcal{L}_{i-1},{P_{i}}} \\ \mathbf{T}_{\mathcal{L}_i/\mathcal{L}_{i-1},{P_{i}}} \end{array}\right]_{\mathcal{R}_{i}}$ that is applied by the link $\mathcal{L}_{i}$ to the substructure $\mathcal{L}_{i-1}$ at point $P_{i}$.
\end{itemize}
and can be represented by the block-diagram depicted in Fig. \ref{TITOP}b.
The way to obtain such a TITOP model $\mathbf{M}_{P_{i}, C_{i}}^{\mathcal{L}_{i}}(s)$ will be detailed later.

\subsection{SDT and Simscape modeling}

\begin{figure}[!ht]
\centering
 \includegraphics[width=1\textwidth]{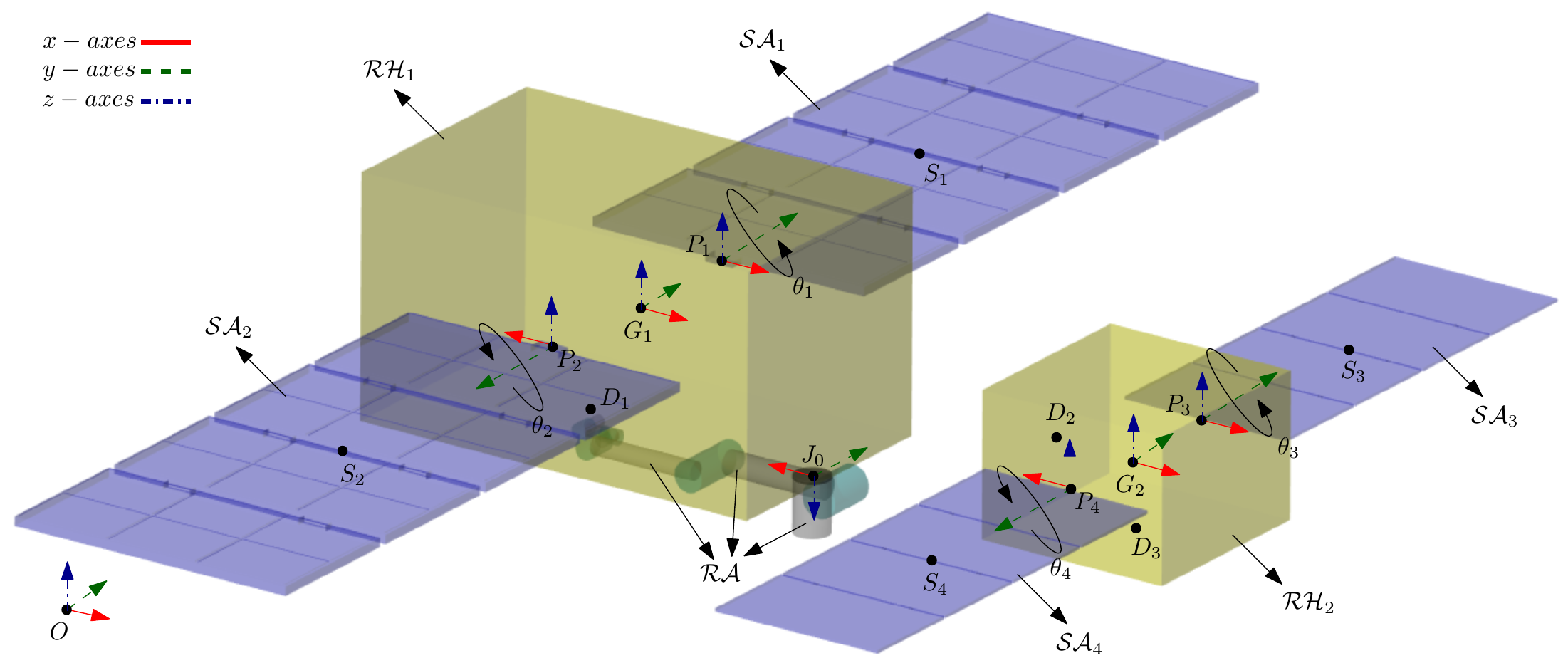}
\caption{Chaser and target spacecraft composed of two rigid hubs, four solar arrays and one robotic arm (Note: for the sake of simplicity, the x-axes are displayed in solid red lines, the y-axes in dashed green lines and the z-axes in dash-dotted blue lines).}
\label{chaser_target_rep} 
\end{figure}

For the on-orbit servicing mission scenario being studied in this paper, two different spacecraft are considered, the chaser and the target, which can be observed in Fig. \ref{chaser_target_rep}. 

\begin{figure}[!ht]
\centering
 \includegraphics[width=1\textwidth]{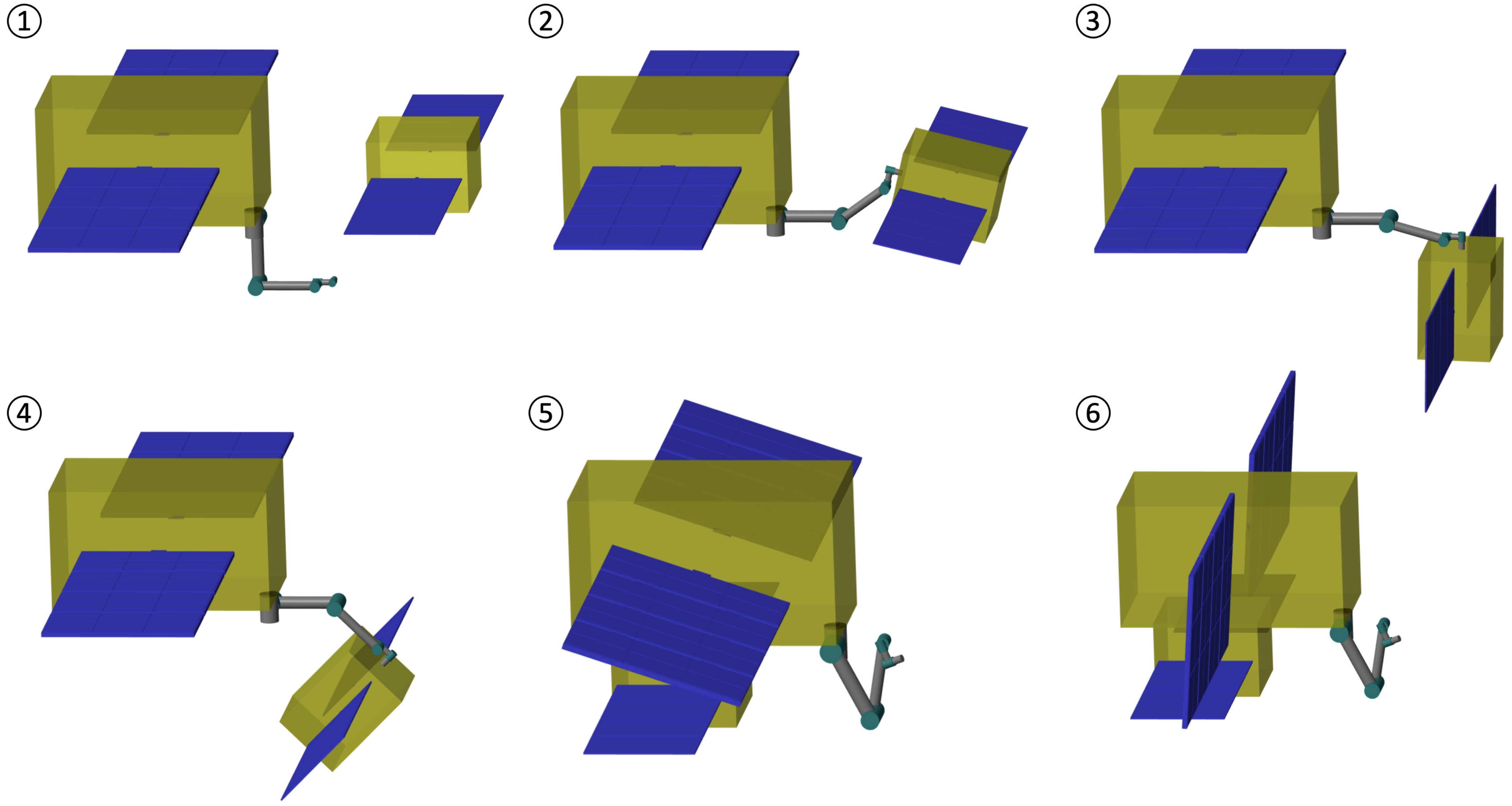}
\caption{Six different illustrations of the decoupled and coupled systems regarding the OOS mission scenario being studied: {\normalsize \textcircled{\scriptsize 1}} decoupled system;  {\normalsize \textcircled{\scriptsize 2}}, {\normalsize \textcircled{\scriptsize 3}} and {\normalsize \textcircled{\scriptsize 4}} the robotic arm has docked to the target spacecraft and it is bringing it closer to the chaser's rigid hub; {\normalsize \textcircled{\scriptsize 5}}, {\normalsize \textcircled{\scriptsize 6}} the target spacecraft is docked to the chaser's rigid hub and the chaser's solar arrays start tilting.}
\label{simscape_rep} 
\end{figure}

The chaser spacecraft is composed of a rigid hub, two symmetric flexible solar arrays and one robotic arm. The target vehicle consists of a rigid hub and two flexible solar arrays. Initially, the chaser's robotic arm is stretched near the bottom surface of the vehicle. While the chaser performs a rendezvous with the target, the arm starts its motion in order to dock to the other spacecraft. After seizing the vehicle, the robotic manipulator attaches it to the chaser's rigid body by means of another docking mechanism. Finally, the chaser's solar panels start rotating with the objective of optimizing the power being provided to the coupled system. For a better understanding of the mission scenario being studied, Fig. \ref{simscape_rep} depicts six different representations of the decoupled and coupled systems during the whole final rendezvous phase. 

\begin{figure}[!ht]
\centering
\includegraphics[width=1\textwidth]{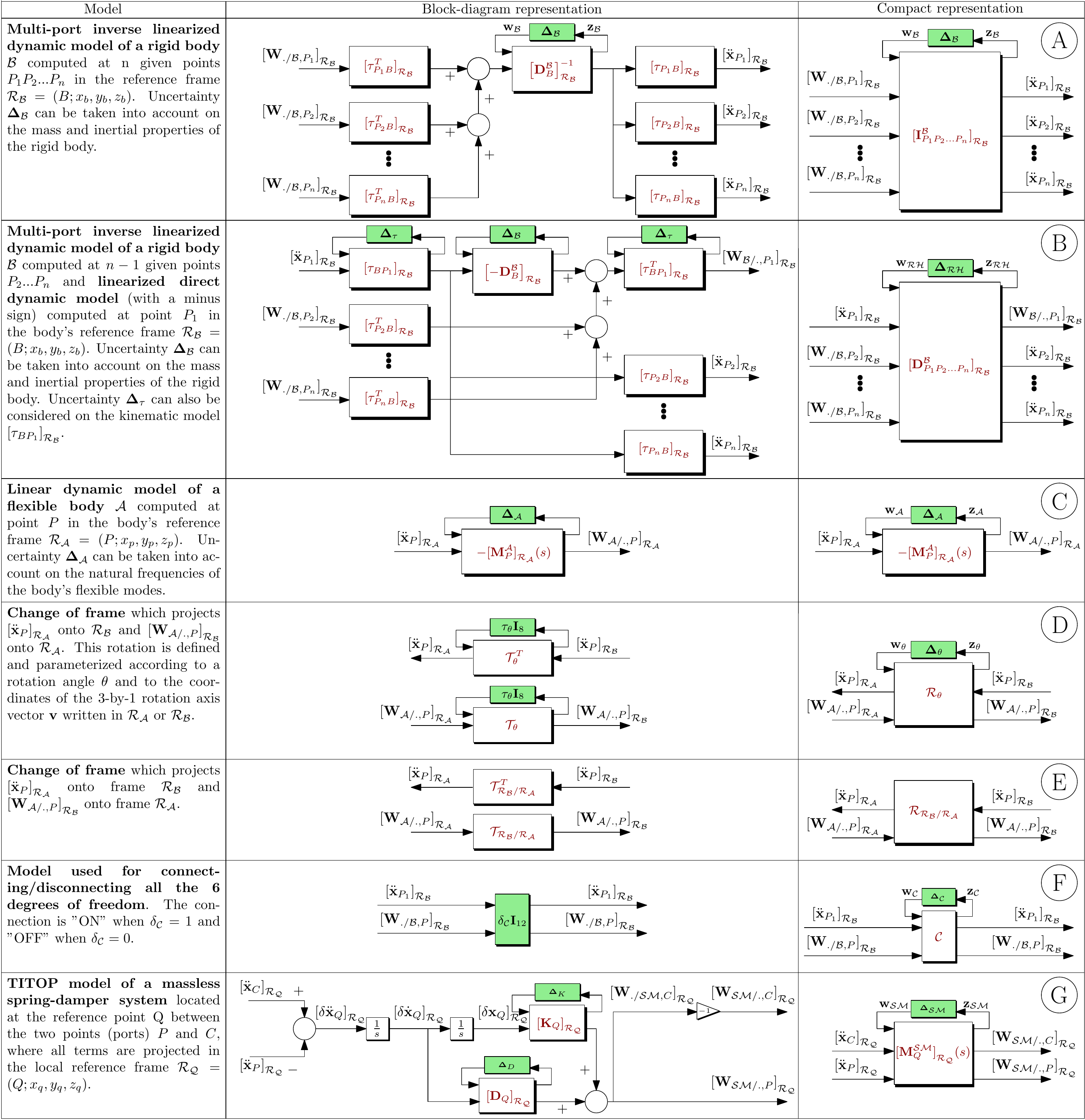}
\captionof{table}[foo]{Block-diagram representations based on the dynamic models of the several elements that are used to build the full model of the system.}
\label{table_models}
\end{figure}

Let us now introduce in Table \ref{table_models} the block-diagram representations based on
the dynamic models of the several elements that will be used to build the full model of the system. A detailed analysis of each model will be provided later in the following sections of this paper. In this table, ${[\tau_{PB}]}_{\mathcal{R}_{\mathcal{B}}}$ describes the rigid kinematic model between the degrees of freedom (DOF) of point $P$ and the DOF of point $B$ projected in frame $\mathcal{R}_{\mathcal{B}}$. It is given by:

\begin{equation}
{[\tau_{P B}]}_{\mathcal{R}_{\mathcal{B}}}=\left[\begin{array}{cc}
\textbf{I}_{3} & \left({ }^{*} \overrightarrow{P B}\right) \\
0_{3 \times 3} & \textbf{I}_{3}
\end{array}\right] \quad \text{with} \quad \left({ }^{*} \overrightarrow{P B}\right)=\left[\begin{array}{ccc}0 & -z & y \\ z & 0 & -x \\ -y & x & 0\end{array}\right]_{\mathcal{R}_{\mathcal{B}}}
\label{kinmodel}
\end{equation}

where $\left({ }^{*} \overrightarrow{PB}\right)$ represents the skew-symmetric matrix that results from the vector $\overrightarrow{P B}$ and $\left[\begin{array}{lll}x & y & z\end{array}\right]_{\mathcal{R}_{\mathcal{B}}}^{T}$ is the coordinate vector of  $\overrightarrow{P  B}$ projected in frame ${\mathcal{R}_{\mathcal{B}}}$. Furthemore, the transformation matrix $\mathcal{T}_{\mathcal{R}_{\mathcal{B}}/\mathcal{R}_{\mathcal{A}}}$ between frames $\mathcal{R}_{\mathcal{B}}$ and $\mathcal{R}_{\mathcal{A}}$ is equal to:

\begin{equation}
\mathcal{T}_{\mathcal{R}_{\mathcal{B}}/\mathcal{R}_{\mathcal{A}}}=\left[\begin{array}{cc}
\mathcal{DCM}_{\mathcal{R}_{\mathcal{B}}/\mathcal{R}_{\mathcal{A}}} & 0_{3 \times 3} \\
0_{3 \times 3} & \mathcal{DCM}_{\mathcal{R}_{\mathcal{B}}/\mathcal{R}_{\mathcal{A}}}
\end{array}\right]
\end{equation}

where $\mathcal{DCM}_{\mathcal{R}_{\mathcal{B}}/\mathcal{R}_{\mathcal{A}}}$ is a Direct Cosine Matrix (DCM) that transforms $\mathcal{R}_{\mathcal{A}}=\left(P; x_{p}, y_{p}, z_{p}\right)$ to $\mathcal{R}_{\mathcal{B}}=\left(B; x_{b}, y_{b}, z_{b}\right)$ (i.e., the matrix of components of unitary vectors $x_{p}, y_{p}, z_{p}$ in $\mathcal{R}_{\mathcal{B}}$). $\mathcal{T}_{\mathcal{R}_{\mathcal{B}}/\mathcal{R}_{\mathcal{A}}}$ is a six-by-six matrix due to the fact that six DOF are being considered (three translations and three rotations).

\subsubsection{Rigid hub model}
A general rigid hub $\mathcal{RH}$ with center of mass $G$ can be modeled using the SDT's block \textit{Multi-port rigid body}, which computes the inverse linearized dynamic model of a rigid body at \textit{n} given points, as explained in the SDT's user guide \cite{userguide}. Considering that the rigid body is submitted to external forces/moments $\mathbf{F}_{\mathrm{ext}}, \mathbf{T}_{\mathrm{ext}, G}$ and to forces/moments
$\mathbf{F}_{\mathcal{RH} / \mathcal{A}}, \mathbf{T}_{\mathcal{RH}/ \mathcal{A}, P}$ due to the interactions with an appendage $\mathcal{A}$ connected at point $P$, the linearized Newton/Euler equations read:

\begin{equation}
\left[\begin{array}{c}\mathbf{F}_{\mathrm{ext}}-\mathbf{F}_{\mathcal{RH} / \mathcal{A}} \\ \mathbf{T}_{\mathrm{ext}, G}-\mathbf{T}_{\mathcal{RH} / \mathcal{A}, G}\end{array}\right]=\textbf{D}_{G}^{\mathcal{RH}}\left[\begin{array}{c}\mathbf{a}_{G} \\ \boldsymbol{\dot{\omega}}\end{array}\right] \quad \text { with } \quad \textbf{D}_{G}^{\mathcal{RH}}=\left[\begin{array}{cc}m^{\mathcal{RH}} \textbf{I}_{3} & 0_{3 \times 3} \\ 0_{3 \times 3} & \textbf{J}_{G}^{\mathcal{RH}}\end{array}\right]
\label{rheq}
\end{equation}

where $\textbf{D}_{G}^{\mathcal{RH}}$ is the static direct dynamic model of $\mathcal{RH}$ at the body's center of mass $G$ and $\mathbf{a}_{G}$ is the linear acceleration vector of $\mathcal{RH}$ at $G$. Furthermore, $m^{\mathcal{RH}}$ is the mass of $\mathcal{RH}$ and $\textbf{J}_{G}^{\mathcal{RH}}$ represents the inertia tensor of $\mathcal{RH}$ written in the body frame of the rigid hub.

Looking now at the case of the OOS mission scenario being studied, the rotating body frames of the rigid hubs of the chaser and target spacecraft are represented in Fig. \ref{chaser_target_rep} (for the sake of simplicity, the x-axes are displayed in solid red lines, the y-axes in dashed green lines and the z-axes in dash-dotted blue lines). The body reference frame of the chaser's rigid hub $\mathcal{RH}_{1}$ is given by $\mathcal{R}_{\mathcal{RH}_{1}}=\left({G_{1}}; x_{G_{1}}, y_{G_{1}}, z_{G_{1}}\right)$, where ${G_{1}}$ is the center of mass and reference point of the main body. The same applies to the target's rigid hub $\mathcal{RH}_{2}$, with $\mathcal{R}_{\mathcal{RH}_{2}}=\left({G_{2}}; x_{G_{2}}, y_{G_{2}}, z_{G_{2}}\right)$.

Since the chaser's rigid hub is considered to be the system's main body, ${\boldsymbol{\dot{\omega}}}$ is denoted as the angular velocity vector of the body frame $\mathcal{R}_{\mathcal{RH}_{1}}$ with respect to the inertial frame $\mathcal{R}_{O}=\left(O; x_{O}, y_{O}, z_{O}\right)$, expressed in $\mathcal{R}_{\mathcal{RH}_{1}}$. Initially, when the chaser spacecraft is decoupled from the target, $\mathcal{RH}_{1}$ is connected to two solar arrays at points ${P}_{1}$ and ${P}_{2}$ as well as to one robotic arm at point ${J}_{0}$. However, point ${D}_{3}$ of $\mathcal{RH}_{2}$ will eventually dock to point ${D}_{1}$ of $\mathcal{RH}_{1}$, as can be observed in illustrations {\Large \textcircled{\normalsize 5}} and {\Large \textcircled{\normalsize 6}} in Fig. \ref{simscape_rep}. For that reason, this port also needs to be considered. Furthermore, it is also considered that external forces and torques $\mathbf{W}_{e x t, G_{1}}=\left[\begin{array}{c}\mathbf{F}_{e x t} \\ \mathbf{T}_{e x t, G_{1}}\end{array}\right]$ are acting on $\mathcal{RH}_{1}$ at point ${G}_{1}$. $\mathcal{RH}_{1}$ is finally modeled using the compact representation {\Large \textcircled{\normalsize A}} shown in Table \ref{table_models}. The inverse linearized dynamic model of $\mathcal{RH}_{1}$ defined in $\mathcal{R}_{\mathcal{RH}_{1}}$ is then equal to ${[\textbf{I}_{G_{1}P_{1}P_{2}J_{0}D_{1}}^{\mathcal{RH}_1}]}_{\mathcal{R}_{\mathcal{RH}_1}}$.

\textbf{Connection between rigid hubs}: The frame transformation block-diagram $\mathcal{R}_{{\mathcal{R}_{\mathcal{RH}_1}}/{\mathcal{R}_{\mathcal{RH}_2}}}$ based on the compact representation {\Large \textcircled{\normalsize E}} in Table \ref{table_models} is used to connect $\mathcal{RH}_1$ to $\mathcal{RH}_2$ at point ${D}_1 \equiv {D}_3$.

Similarly to the chaser's rigid hub, $\mathcal{RH}_{2}$ is connected to two solar arrays at points ${P}_{3}$ and ${P}_{4}$. However, two docking phases will occur. Firstly, the end effector of the robotic arm located at point $J_{7}$ will dock to point $D_{2}$ of the target's rigid hub $\mathcal{RH}_{2}$, as can be seen in illustrations {\Large \textcircled{\normalsize 5}} and {\Large \textcircled{\normalsize 6}}, with point $J_{7}$ being displayed in Fig. \ref{roboticarm}a. Secondly, port ${D}_{3}$ of $\mathcal{RH}_{2}$ will also dock to port ${D}_{1}$ of $\mathcal{RH}_{1}$, as referred before.

\textbf{Mechanical uncertainty} $\boldsymbol{\Delta}_{mec}$: One of the objectives of this paper is to demonstrate how to design a controller in the presence of significant model uncertainty. For that reason, relative uncertainty is taken into account on the mass and moments of inertia of $\mathcal{RH}_{2}$, since the robust stability and performance of the coupled system need to be ensured even when the mechanical characteristics of the target's rigid hub are not perfectly known. As an example, let us now consider the mass of the target's rigid body $m^{\mathcal{RH}_{2}}$ as uncertain:

\begin{equation}
m^{\mathcal{RH}_{2}}=m^{\mathcal{RH}_{2}}_{0}(1+r_{m^{\mathcal{RH}_{2}}}\delta_{m^{\mathcal{RH}_{2}}}) 
\label{massunc}
\end{equation}

where $m^{\mathcal{RH}_{2}}_{0}$ is the body's nominal mass, $r_{m^{\mathcal{RH}_{2}}}$ is used to set the maximum percent of variation for the body's mass and $\delta_{m^{\mathcal{RH}_{2}}}  \in [-1, 1]$ is a normalized real uncertainty. In the case of the mass, $\delta_{m^{\mathcal{RH}_{2}}}$ appears three times in a minimal LFR of a rigid body \cite{userguide}. Therefore, the uncertainty block regarding the mass of $\mathcal{RH}_{2}$ is equal to $\boldsymbol{\Delta}_{m^{\mathcal{RH}_{2}}}=\delta_{m^{\mathcal{RH}_{2}}}\textbf{I}_{3}$. Similarly, relative uncertainty is also considered on the inertial properties of ${\mathcal{RH}_{2}}$. However, only the diagonal moments of inertia were assumed to be uncertain while the off-diagonal terms are kept at their nominal values. For the moments of inertia of the target's rigid hub $\textbf{J}_{xx_{\mathcal{RH}_{2}}}$, $\textbf{J}_{yy_{\mathcal{RH}_{2}}}$ and $\textbf{J}_{zz_{\mathcal{RH}_{2}}}$, the normalized real uncertainties $\delta_{\textbf{J}_{xx_{\mathcal{RH}_{2}}}}$, $\delta_{\textbf{J}_{yy_{\mathcal{RH}_{2}}}}$ and $\delta_{\textbf{J}_{zz_{\mathcal{RH}_{2}}}}$ have just one occurrence in the same minimal LFR \cite{userguide}, where $\delta_{\textbf{J}_{\bullet_{\mathcal{RH}_{2}}}} \in [-1, 1]$ and $r_{\textbf{J}_{\bullet_{\mathcal{RH}_{2}}}}$ are used to set the maximum percent of variation for $\textbf{J}_{\bullet_{\mathcal{RH}_{2}}}$, just like in Eq. (\ref{massunc}). Therefore, the mechanical uncertainty block of the target's rigid body can be written as $\boldsymbol{\Delta}_{mec}=\operatorname{diag}\left(\boldsymbol{\Delta}_{m^{\mathcal{RH}_{2}}}, \delta_{\textbf{J}_{xx_{\mathcal{RH}_{2}}}}, \delta_{\textbf{J}_{yy_{\mathcal{RH}_{2}}}}, \delta_{\textbf{J}_{zz_{\mathcal{RH}_{2}}}}\right)$. The dynamic model of $\mathcal{RH}_2$ will although be defined later in this paper.

In \textit{Simscape}, the \textit{Inertia} block is used to model a rigid body, which assumes the mass to be distributed in space, allowing the rigid hubs to have non-zero moments of inertia, products of inertia and center-of-mass coordinates.

\subsubsection{Dynamic model of a cantilevered solar array}

A general flexible solar array $\mathcal{SA}$ connected to a parent body $\mathcal{RH}$ at point $P$ can be modeled in SDT using the block \textit{One-port flexible body}, as explained in \cite{userguide}. The effective mass model of the solar array $\mathbf{M}_{P}^{\mathcal{SA}}(s)$ relates the six DOF acceleration vector of point $P$ and the six DOF forces/moments vector applied by the parent body to the appendage $\mathcal{SA}$ at point $P$:

\begin{equation}
\left[\begin{array}{c}\mathbf{F}_{\mathcal{RH} / \mathcal{SA}} \\ \mathbf{T}_{\mathcal{RH} / \mathcal{SA}, P}\end{array}\right]=\mathbf{M}_{P}^{\mathcal{SA}}(s)\left[\begin{array}{c}\mathbf{a}_{P} \\ \boldsymbol{\dot{\omega}}\end{array}\right] \quad \text { with } \quad \mathbf{M}_{P}^{\mathcal{SA}}(s)=\mathbf{D}_{P, 0}^{\mathcal{SA}}+\Sigma_{i=1}^{N} \mathbf{M}_{i, P}^{\mathcal{SA}} \frac{2 \xi_{i_\mathcal{{S\!A}}}\omega_{i_\mathcal{{S\!A}}} s+\omega_{i_\mathcal{{S\!A}}}^{2}}{s^{2}+2 \xi_{i_\mathcal{{S\!A}}} \omega_{i_\mathcal{{S\!A}}} s+\omega_{i_\mathcal{{S\!A}}}^{2}} 
\label{sadynmodel}
\end{equation}

where:

\begin{itemize}
    \item $\mathbf{a}_{P}$ is the linear acceleration vector of $\mathcal{SA}$ at $P$.
    \item $\omega_{i_\mathcal{{S\!A}}}, \xi_{i_\mathcal{{S\!A}}}$ and $\boldsymbol{l}_{i, P}^{\mathcal{SA}}$ are the natural frequency, the damping ratio and the 6 DOF participation factor vector of the $i$-th flexible mode of the appendage $\mathcal{SA}$.
    \item $\mathbf{L}_{P}^{\mathcal{SA}}=\left[\boldsymbol{l}_{1, P}^{\mathcal{SA}}, \ldots, \boldsymbol{l}_{i, P}^{\mathcal{SA}}, \ldots, \boldsymbol{l}_{N, P}^{\mathcal{SA}}\right]$ is the matrix of modal participation factors of the $N$ flexible modes of the appendage at point $P$.    
    \item $\mathbf{D}_{P,0}^{\mathcal{SA}}=\mathbf{D}_{P}^{\mathcal{SA}}-\Sigma_{i=1}^{N} \mathbf{M}_{i,P}^{\mathcal{SA}}=\mathbf{D}_{P}^{\mathcal{SA}}-\mathbf{L}_{P}^{\mathcal{SA}} {\mathbf{L}_{P}^{\mathcal{SA}}}^{T}$ is the 6-by-6 residual mass/inertia of the appendage rigidly cantilevered to the parent body $\mathcal{RH}$ at point $P$.
    \item $\mathbf{M}_{i, P}^{\mathcal{SA}}=\boldsymbol{l}_{i, P}^{\mathcal{SA}} {\boldsymbol{l}_{i, P}^{\mathcal{SA}}}^{T}$ is the 6-by-6 effective mass/inertia matrix of the $i$-th flexible mode of the appendage. 

\end{itemize}{}

MATLAB’s Partial Differential Equation (PDE) Toolbox is used to perform finite element analysis and extract all these structural dynamics parameters based on the 3D model of each solar array, its material properties and the boundary conditions. All the reference frames of the solar arrays $\mathcal{R}_{\mathcal{SA}_{\bullet}}=\left({P_{\bullet}}; x_{P_{\bullet}}, y_{P_{\bullet}}, z_{P_{\bullet}}\right)$ are depicted in Fig. \ref{chaser_target_rep}, where points $P_{\bullet}$ represent the connection and reference points of the four flexible appendages. In addition, points $S_{\bullet}$ are the solar arrays' centers of mass. The effective mass models of the solar arrays are based on the compact representation {\Large \textcircled{\normalsize C}} shown in Table \ref{table_models} and equal to $-{[\textbf{M}_{P_{\bullet}}^{\mathcal{SA}_{\bullet}}]}_{\mathcal{R}_{\mathcal{SA}_{\bullet}}}(s)$, with $\mathbf{M}_{P_{\bullet}}^{\mathcal{SA}_{\bullet}}(s)$ being described in Eq. (\ref{sadynmodel}).

In \textit{Simscape}, the same process is achieved by means of the \textit{Reduced Order Flexible Solid} block, which takes into account three mass, stiffness and damping matrices which are obtained by applying the Craig-Bampton order reduction method while considering the solar arrays $\mathcal{R}_{\mathcal{SA}_{\bullet}}$ to be cantilevered at points $P_{\bullet}$. 

\textbf{Modal uncertainty $\boldsymbol{\Delta}_{mod}$}: Changes in structural parameters can provoke variations in the natural frequency of some flexible modes. Since the final goal is to design a controller where the bandwidth of interest can be highly impacted by these parameters, relative uncertainty is taken into consideration on the natural frequencies of all the solar arrays' first flexible modes, which are given by $\omega_{1_\mathcal{{S\!A}_{\bullet}}}$. Similarly to how $\boldsymbol{\Delta}_{mec}$ was constructed, it results that:

\begin{equation}
\omega_{1_\mathcal{{S\!A}_{\bullet}}}=\omega^{0}_{1_\mathcal{{S\!A}_{\bullet}}}(1+r_{\omega_{1_\mathcal{{S\!A}_{\bullet}}}}\delta_{\omega_{1_\mathcal{{S\!A}_{\bullet}}}}) \end{equation}

where $\omega^{0}_{1_\mathcal{{S\!A}_{\bullet}}}$ represents the nominal natural frequencies, $\delta_{\omega_{1_\mathcal{{S\!A}_{\bullet}}}} \in [-1, 1]$ are normalized real uncertainties and the parameters $r_{\omega_{1_\mathcal{{S\!A}_{\bullet}}}}$ are used to set the maximum percent of variation for $\omega_{1_\mathcal{{S\!A}_{\bullet}}}$. Since the uncertainties $\delta_{\omega_{1_{\mathcal{{S\!A}}_{\bullet}}}}$ appear two times per flexible mode in a minimal LFR of a flexible appendage, as explained in \cite{Guy2014}, the modal uncertainty block linked to each solar array $\mathcal{{S\!A}_{\bullet}}$ is equal to $\boldsymbol{\Delta}_{\omega_{\bullet}}=\delta_{\omega_{1_\mathcal{{S\!A}_{\bullet}}}}\textbf{I}_{2}$. Furthermore, the uncertainty block regarding the modal properties of all the four solar arrays can be described as $\boldsymbol{\Delta}_{mod}=\operatorname{diag}\left(\boldsymbol{\Delta}_{\omega_{1}}, \boldsymbol{\Delta}_{\omega_{2}}, \boldsymbol{\Delta}_{\omega_{3}}, \boldsymbol{\Delta}_{\omega_{4}}\right)$.

\textbf{Connection between rigid hubs and solar arrays}: The frame transformation blocks $\mathcal{R}_{{\mathcal{R}_{\mathcal{RH}_{\bullet}}}/{\mathcal{R}^0_{\mathcal{SA}_{\bullet}}}}$ are needed to project  ${[\ddot{\mathbf{x}}_{P_{\bullet}}]}_{\mathcal{R}_{\mathcal{RH}_{\bullet}}}$ onto frames ${\mathcal{R}^0_{\mathcal{SA}_{\bullet}}}$ and ${[\mathbf{W}_{\mathcal{SA}_{\bullet}/\mathcal{RH}_{\bullet},P_{\bullet}}]}_{\mathcal{R}_{\mathcal{R}^0_{\mathcal{SA}_{\bullet}}}}$ onto frames ${\mathcal{R}_{\mathcal{RH}_{\bullet}}}$. Here, frames ${\mathcal{R}^0_{\mathcal{SA}_{\bullet}}}=\left({P_{\bullet}}; x^0_{P_{\bullet}}, y^0_{P_{\bullet}}, z^0_{P_{\bullet}}\right)$ correspond exactly to frames ${\mathcal{R}_{\mathcal{SA}_{\bullet}}}$ when the geometrical configuration of the solar arrays $\theta_{\bullet}$ is equal to 0 $rad$. These blocks are based on the compact representation defined in {\Large \textcircled{\normalsize E}}.

 \textbf{Varying tilt angles of the solar arrays} $\theta_{\bullet}$: The system is also parameterized according to the solar arrays' tilt angles $\theta_{\bullet}$, respectively expressed in the reference frames $\mathcal{R}_{\mathcal{SA}_{\bullet}}$. The fact that these configurations are considered as time-varying is of paramount importance because these structures are cantilevered on the main bodies of both spacecraft with a varying tilt angle, which has a direct influence on the dynamic behavior of both the decoupled and coupled systems. In essence, the attitude control system must be robust to such a variation. The frame transformation blocks $\mathcal{R}_{\theta_{\bullet}}$ defined in {\Large \textcircled{\normalsize D}} are  needed to project  ${[\ddot{\mathbf{x}}_{P_{\bullet}}]}_{\mathcal{R}^0_{\mathcal{SA}_{\bullet}}}$ onto frames ${\mathcal{R}_{\mathcal{SA}_{\bullet}}}$ and ${[\mathbf{W}_{\mathcal{SA}_{\bullet}/\mathcal{RH}_{\bullet},P_{\bullet}}]}_{\mathcal{R}_{\mathcal{R}_{\mathcal{SA}_{\bullet}}}}$ onto frames ${\mathcal{R}^0_{\mathcal{SA}_{\bullet}}}$. These block-diagrams are parameterized according to $\tau_{\theta_{\bullet}}=\tan(\theta_{\bullet}/4)$, which leads to a minimal LFR-type representation, as proposed by Dubanchet in \cite{dubanchetphd}. Since $\tau_{\theta_{\bullet}}$ is repeated eight times per $\mathcal{T}_{\theta_{\bullet}}$, $\tau_{\theta_{\bullet}}$ appears sixteen times for each connection between a rigid hub and a solar array. In the end, the uncertainty block describing the varying tilt angles of the flexible appendages is equal to $\boldsymbol{\Delta}_{\theta}=\operatorname{diag}\left(\boldsymbol{\Delta}_{\theta_{1}}, \boldsymbol{\Delta}_{\theta_{2}}, \boldsymbol{\Delta}_{\theta_{3}}, \boldsymbol{\Delta}_{\theta_{4}}\right)$, with $\boldsymbol{\Delta}_{\theta_{\bullet}}=\tau_{\theta_{\bullet}}\textbf{I}_{16}$ and $\tau_{\theta_{\bullet}} \in [-1, 1]$ (which characterizes a complete revolution of the flexible appendages). It should also be noted that $\theta_{\bullet}=0$ $rad$ in Fig. \ref{chaser_target_rep}. $\mathcal{R}_{\theta_{\bullet}}$ are also defined according to the coordinates of 3-by-1 rotation axis vectors. In this case, the rotations happen around $y_{P_{\bullet}} \equiv y^0_{P_{\bullet}}$, as depicted in Fig. \ref{chaser_target_rep}. 
 
 In \textit{Simscape}, the same process is achieved by attaching \textit{Revolute Joint} blocks to the respective \textit{Reduced Order Flexible Solid} blocks. Furthermore, the revolute joints are considered to be actuated in motion, meaning that the physical signal input provides the desired trajectory and the actuation torque is automatically computed and applied based on model dynamics.

\subsubsection{Robotic arm model}
\begin{figure}[!ht]
\centering
 \includegraphics[width=1\textwidth]{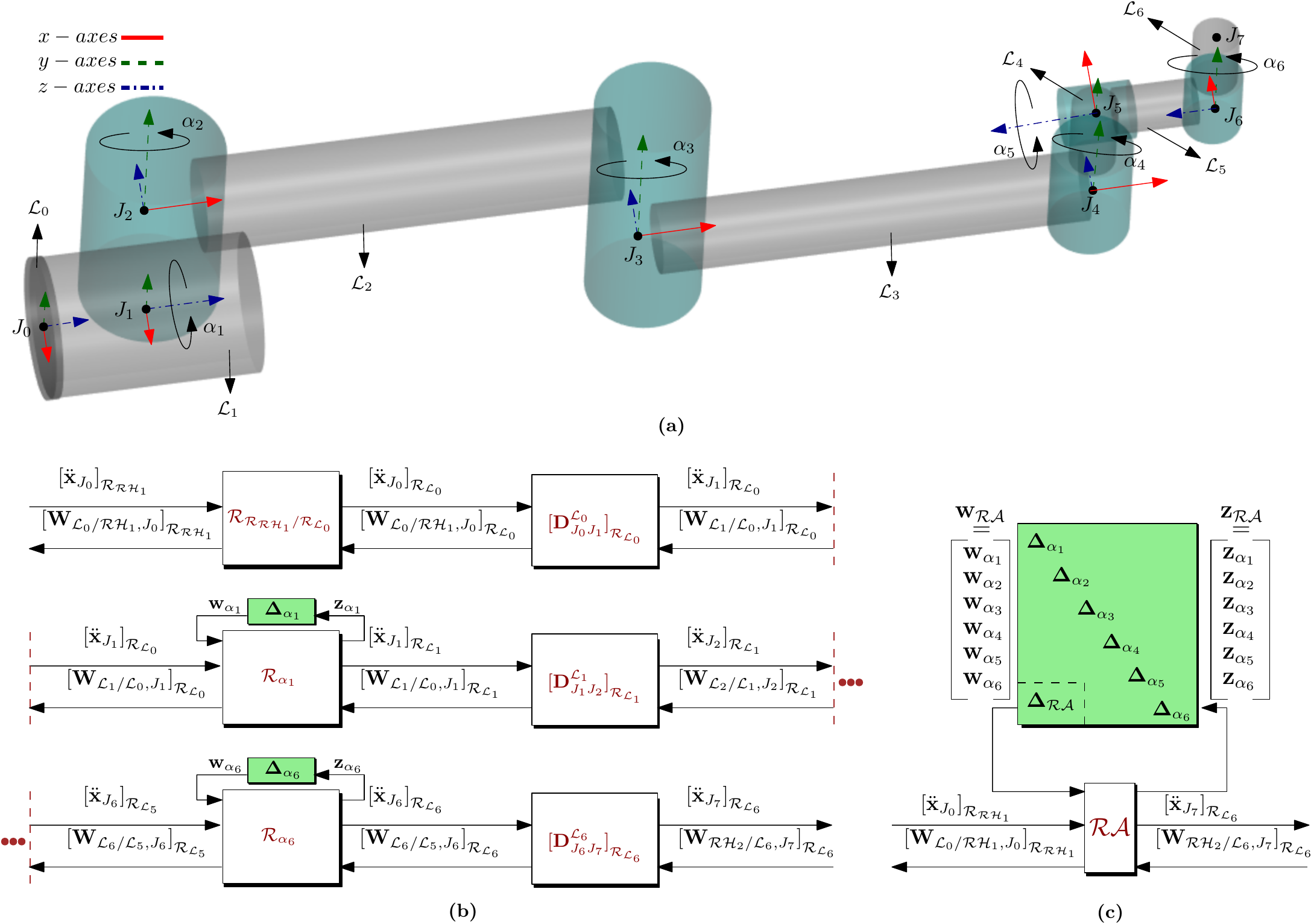}
\caption{Robotic manipulator representation: (a) robotic arm kinematics (Note: for the sake of simplicity, the x-axes are displayed in solid red lines, the y-axes in dashed green lines and the z-axes in dash-dotted blue lines). (b) block-diagram of the parameterized robotic arm written in LFR form. (c) equivalent LFR form of the manipulator.}
\label{roboticarm} 
\end{figure}

Ultimately, the chaser spacecraft uses a robotic arm $\mathcal{RA}$ for catching the target on-orbit, which is inspired by the Universal Robots' UR5 robotic arm \cite{Kebria2017}. First, it should be noted that $\alpha_{\bullet}$ represents the arm's angular configuration, as depicted in Fig. \ref{roboticarm}a. In addition, all the arm's 7 different links $\mathcal{L}_{\bullet}$ are assumed to be rigid. Each link $\mathcal{L}_{\bullet}$ can be taken into account using the \textit{Multi-port rigid body} block of the SDT library. The TITOP dynamic model of each link $\mathcal{L}_{i}$ (for $i=0...6$) is given by ${[\textbf{D}_{J_{i}J_{i+1}}^{\mathcal{L}_i}]}_{\mathcal{R}_{\mathcal{L}_i}}$, since  $\mathcal{L}_{i}$ is connected to a parent substructure at point $J_{i}$ and to a child substructure at point $J_{i+1}$. 

\textbf{Connection between the chaser's rigid hub and the robotic arm}: The frame transformation block-diagram $\mathcal{R}_{{\mathcal{R}_{\mathcal{RH}_1}}/{\mathcal{R}_{\mathcal{L}_0}}}$ is used to connect the chaser's rigid hub to the robotic arm $\mathcal{RA}$ at point ${J}_0$.

\textbf{Varying tilt angles of the robotic arm} $\alpha_{\bullet}$: The robotic manipulator is assumed to be motion actuated. The system is also parameterized according to the manipulator's geometrical configuration $\alpha_{\bullet}$, respectively expressed in the reference frames $\mathcal{R}_{\mathcal{L}_{\bullet}}=\left({J_{\bullet}}; x_{J_{\bullet}}, y_{J_{\bullet}}, z_{J_{\bullet}}\right)$ displayed in Fig. \ref{roboticarm}a. This means that the block-diagram depicted in  {\Large \textcircled{\normalsize E}} is once again used to model the change of frame blocks $\mathcal{R}_{\alpha_{\bullet}}$, as can be observed in the block-diagram displayed in Fig. \ref{roboticarm}b. Similarly to how $\boldsymbol{\Delta}_{\theta}$ was built, the uncertainty block describing the changing geometrical configuration of the robotic arm is given by $\boldsymbol{\Delta}_{\mathcal{RA}}=\operatorname{diag}\left(\boldsymbol{\Delta}_{\alpha_{1}}, \boldsymbol{\Delta}_{\alpha_{2}}, \boldsymbol{\Delta}_{\alpha_{3}}, \boldsymbol{\Delta}_{\alpha_{4}},
\boldsymbol{\Delta}_{\alpha_{5}},
\boldsymbol{\Delta}_{\alpha_{6}}\right)$, with $\boldsymbol{\Delta}_{\alpha_{\bullet}}=\tau_{\alpha_{\bullet}}I_{16}$ and $\tau_{\alpha_{\bullet}} \in [-1, 1]$. Fig. \ref{roboticarm}c displays the equivalent global LFR form of the robotic arm, where $\mathbf{w}_{\mathcal{RA}}\boldsymbol{\Delta}_{\mathcal{RA}}=\mathbf{z}_{\mathcal{RA}}$. Furthermore, $\mathbf{w}_{\mathcal{RA}}$ and $\mathbf{z}_{\mathcal{RA}}$ are the endogenous inputs and outputs of the arm manipulator model. It should also be noted that the angular configuration of the arm represented in Fig. \ref{roboticarm}a is given by $\alpha_{2,4}=-\pi/2$ $rad$ and $\alpha_{1,3,5,6}=0$ $rad$. 

In \textit{Simscape}, the robotic arm model consists of a combination of \textit{Inertia} and \textit{Revolute Joint} blocks.

\subsubsection{Modeling of the decoupled/coupled configurations}

The coupled and decoupled configurations are modeled with the help of two real parametric uncertainties $\delta_{\mathcal{C}_{1}}$ and $\delta_{\mathcal{C}_{2}}$, as described in {\Large \textcircled{\normalsize F}}. If $\delta_{\mathcal{C}_{\bullet}}=0$, the respective channels are completely switched off and there is no attachment between bodies. In case $\delta_{\mathcal{C}_{\bullet}}=1$, then docking has occurred and the bodies are connected to each other. Three different possibilities are considered. The first one can be seen in representation {\Large \textcircled{\normalsize 1}}, where the target is completely detached from the chaser. In this case, both $\delta_{\mathcal{C}_{1}}$ and $\delta_{\mathcal{C}_{2}}$ are equal to 0. Moreover, the second case is illustrated in representations {\Large \textcircled{\normalsize 2}}, {\Large \textcircled{\normalsize 3}} and {\Large \textcircled{\normalsize 4}}, where $\delta_{\mathcal{C}_{1}}=1$ and  $\delta_{\mathcal{C}_{2}}=0$, since the robotic arm is docked to the target. The third and last case occurs when the target is attached to the chaser's rigid hub and completely disconnected from the robotic arm, as can be observed in {\Large \textcircled{\normalsize 5}} and {\Large \textcircled{\normalsize 6}}. In this case, $\delta_{\mathcal{C}_{1}}=0$ and  $\delta_{\mathcal{C}_{2}}=1$. Since the objective is to connect/disconnect all the six DOF, the uncertainty blocks that allow for the modeling of the decoupled/coupled configurations in a single LFR are equal to $\boldsymbol{\Delta}_{\mathcal{C}_{1}}=\delta_{\mathcal{C}_{1}}\textbf{I}_{12}$ and $\boldsymbol{\Delta}_{\mathcal{C}_{2}}=\delta_{\mathcal{C}_{2}}\textbf{I}_{12}$. 

All the required information to build the dynamic model of $\mathcal{RH}_2$ has now been obtained. One of the inputs of the model ${[\textbf{I}_{G_{1}P_{1}P_{2}J_{0}D_{1}}^{\mathcal{RH}_1}]}_{\mathcal{R}_{\mathcal{RH}_1}}$
is the wrench applied by $\mathcal{RH}_{2}$ to $\mathcal{RH}_{1}$ when both bodies are attached at point ${D}_{1} \equiv {D}_{3}$, which has necessarily to be one of the outputs of the dynamic model of $\mathcal{RH}_{2}$. Similarly, ${[\mathbf{W}_{\mathcal{RH}_2/\mathcal{L}_6,J_7}]}_{\mathcal{R}_{\mathcal{L}_6}}$ is one of the robotic arm model inputs when ${J}_{7} \equiv {D}_{2}$. Therefore, the linearized inverse dynamic model of $\mathcal{RH}_{2}$ has to be computed for points ${P}_{3}$, ${P}_{4}$ and the linearized direct dynamic model (with a minus sign) of $\mathcal{RH}_{2}$ has to be computed for points ${D}_{2}$, ${D}_{3}$. However, only one port can be inverted, which means points ${D}_{2}$ and ${D}_{3}$ have to share the same port. The result is the model ${[\textbf{D}_{D_{2/3}P_{3}P_{4}}^{\mathcal{RH}_2}]}_{\mathcal{R}_{\mathcal{RH}_2}}$ defined in $\mathcal{R}_{\mathcal{RH}_{2}}$, which is achieved with the block-diagram {\Large \textcircled{\normalsize B}} shown in Table \ref{table_models}. As stated before, the uncertainty block $\boldsymbol{\Delta}_{mec}$ is taken into account on the linearized model of $\mathcal{RH}_{2}$. Furthermore, the SDT's block \textit{Multi-port rigid body} takes as input the positions of the connection points of $\mathcal{RH}_{2}$ defined in ${\mathcal{R}_{\mathcal{RH}_2}}$ with respect to the reference point $G_{2}$. Since ${D}_{2}$ and ${D}_{3}$ share the same port, the position of the connection point ${D}_{2}$/${D}_{3}$ (inverted port) with respect to $G_2$ is given by $\delta_{\mathcal{C}_{1}}\left(\overrightarrow{G_{2} D_{2}}\right) + \delta_{\mathcal{C}_{2}}\left(\overrightarrow{G_{2} D_{3}}\right)$, considering that both $\delta_{\mathcal{C}_{1}}$ and $\delta_{\mathcal{C}_{2}}$ cannot be equal to 1 at the same time. From Eq. (\ref{kinmodel}), the corresponding skew-symmetric matrix is equal to: 

\begin{equation}
    \left({ }^{*} \overrightarrow{G_{2} {D}_{2}/{D}_{3}}\right)=\left[\begin{array}{ccc}0 & -\delta_{\mathcal{C}_{1}}z_{G_{2} D_{2}}-\delta_{\mathcal{C}_{2}}z_{G_{2} D_{3}} & \delta_{\mathcal{C}_{1}}y_{G_{2} D_{2}}+ \delta_{\mathcal{C}_{2}}y_{G_{2} D_{3}} \\ \delta_{\mathcal{C}_{1}}z_{G_{2} D_{2}}+ \delta_{\mathcal{C}_{2}}z_{G_{2} D_{3}} & 0 & -\delta_{\mathcal{C}_{1}}x_{G_{2} D_{2}}-\delta_{\mathcal{C}_{2}}x_{G_{2} D_{3}} \\ -\delta_{\mathcal{C}_{1}}y_{G_{2} D_{2}}-\delta_{\mathcal{C}_{2}}y_{G_{2} D_{3}} & \delta_{\mathcal{C}_{1}}x_{G_{2} D_{2}}+ \delta_{\mathcal{C}_{2}}x_{G_{2} D_{3}} & 0\end{array}\right]_{\mathcal{R}_{\mathcal{RH}_2}}
\end{equation}

with $\overrightarrow{G_2  D_2}=\left[\begin{array}{lll}x_{G_2  D_2} & y_{G_2  D_2} & z_{G_2  D_2}\end{array}\right]_{\mathcal{R}_{\mathcal{RH}_{2}}}^{T}$ and $\overrightarrow{G_2  D_3}=\left[\begin{array}{lll}x_{G_2  D_3} & y_{G_2  D_3} & z_{G_2  D_3}\end{array}\right]_{\mathcal{R}_{\mathcal{RH}_{2}}}^{T}$. Since the rank of the 3-by-3 matrix $\left({ }^{*} \overrightarrow{G_{2} {D}_{2}/{D}_{3}}\right)$ is always equal to 2, each one of the uncertainties $\delta_{\mathcal{C}_{1}}$ and $\delta_{\mathcal{C}_{2}}$ will have 2 occurrences per kinematic model ${[\tau_{G_{2} {D}_{2}/{D}_{3}}]}_{\mathcal{R}_{\mathcal{RH}_2}}$. Looking now at the block-diagram {\Large \textcircled{\normalsize B}} displayed in Table \ref{table_models}, the kinematic model ${[\tau_{G_{2} {D}_{2}/{D}_{3}}]}_{\mathcal{R}_{\mathcal{RH}_2}}$ appears once in the non-transposed form and once in the transposed form. Therefore, the new uncertainty block linked to the body $\mathcal{RH}_{2}$ is equal to $\boldsymbol{\Delta}_{\mathcal{RH}_{2}}=\operatorname{diag}\left(\boldsymbol{\Delta}_{mec}, \delta_{\mathcal{C}_{1}}\textbf{I}_{4}, \delta_{\mathcal{C}_{2}}\textbf{I}_{4}\right)$.

All the numerical values and range of variations of the numerous system parameters which are employed in this section are described in Table \ref{tab:Sat_prop}.

\subsubsection{Complete model of the system}

A global LFR representation is obtained, minimal in terms of mechanical parameter occurrences. This model fully captures the dynamics and interactions between all the subsystems of the OOS scenario being studied: robotic arm, flexible appendages and decoupled/coupled configurations. Furthermore, it also takes into account the various uncertainty effects in a very compact representation.

\begin{figure}[!ht]
\centering
 \includegraphics[width=1\textwidth]{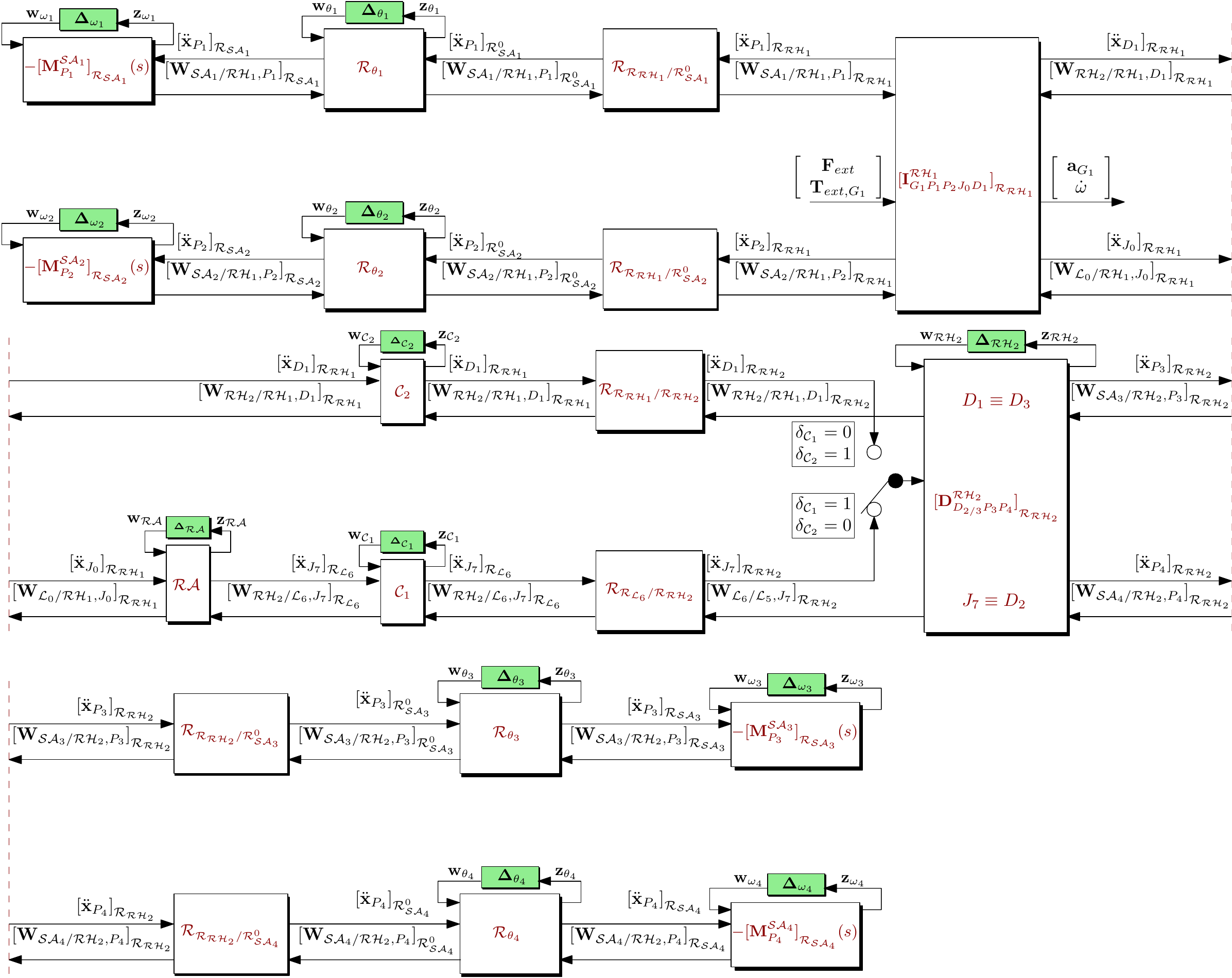}
\caption{Block-diagram of the uncertain plant written in LFR form.}
\label{complete_lft} 
\end{figure}

\begin{figure}[!ht]
\centering
 \includegraphics[width=0.4\textwidth]{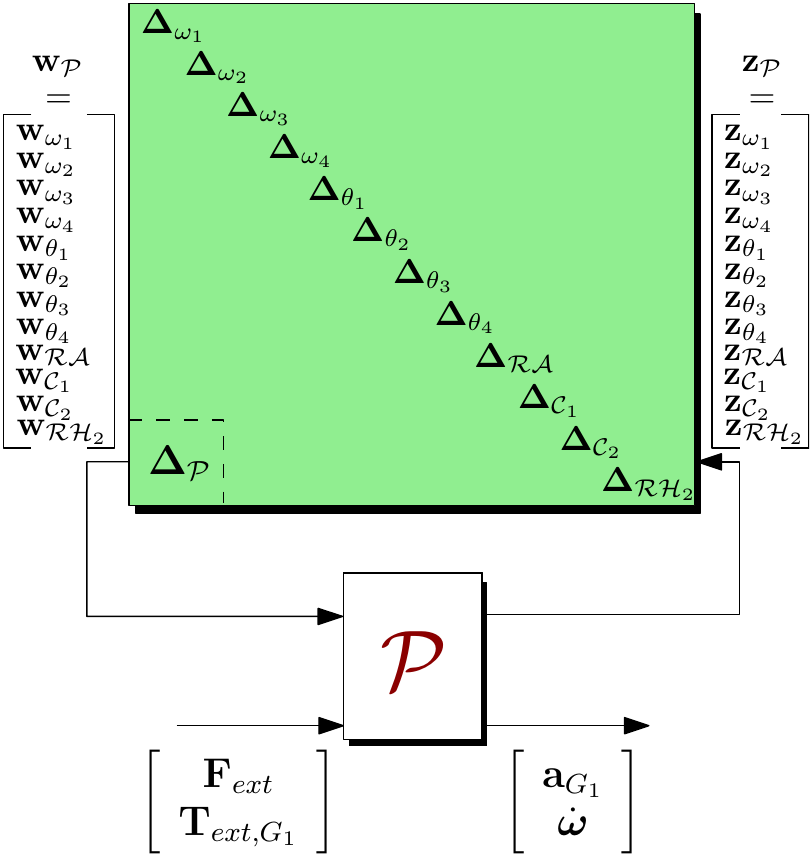}
\caption{Equivalent global LFR form of the complete system.}
\label{compact_lft} 
\end{figure}

Fig. \ref{complete_lft} illustrates the internal structure of the overall LFR model as well as the interconnections between the several subsystems. In this representation, all the block uncertainties are isolated at the component level. However, a low order global uncertainty block system can be built in a very straightforward way by just concatenating the individual uncertainty blocks \cite{Preda2020}, as shown in Fig. \ref{compact_lft}, with $\boldsymbol{\Delta}_{\mathcal{P}}=\operatorname{diag}\left(\boldsymbol{\Delta}_{\omega_1}, \boldsymbol{\Delta}_{\omega_2}, \boldsymbol{\Delta}_{\omega_3}, \boldsymbol{\Delta}_{\omega_4},
\boldsymbol{\Delta}_{\theta_1}, \boldsymbol{\Delta}_{\theta_2}, \boldsymbol{\Delta}_{\theta_3}, \boldsymbol{\Delta}_{\theta_4},
\boldsymbol{\Delta}_{\mathcal{RA}}, \boldsymbol{\Delta}_{\mathcal{C}_1}, \boldsymbol{\Delta}_{\mathcal{C}_2}, \boldsymbol{\Delta}_{\mathcal{RH}_2}\right)$ and $\mathbf{w}_{\mathcal{P}}\boldsymbol{\Delta}_{\mathcal{P}}=\mathbf{z}_{\mathcal{P}}$. Nevertheless, this LFR does not allow for the investigation of what occurs when docking takes place. For that reason, spring-damper systems are introduced.

\subsubsection{Connection model using two spring-damper systems}

\begin{figure}[!ht]
\centering
\includegraphics[width=0.9\textwidth]{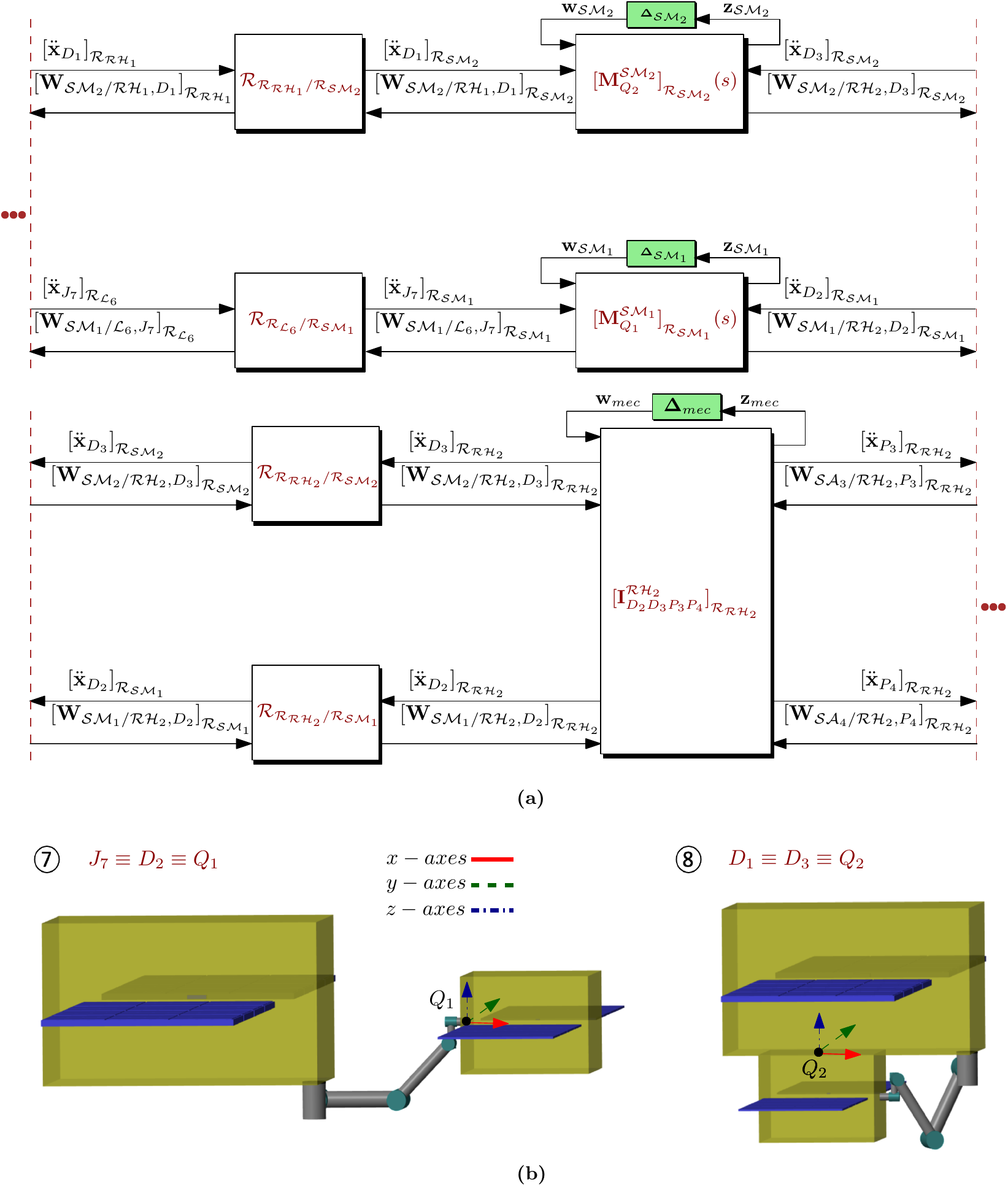}
\caption{(a) Connection model considering two spring-damper systems. (b) Two different moments where docking takes place: {\normalsize \textcircled{\scriptsize 7}} the robotic arm docks to the target spacecraft; {\normalsize \textcircled{\scriptsize 8}} the target's rigid hub docks to the chaser's rigid hub and the robotic arm disconnects from the target afterwards (Note: for the sake of simplicity, the x-axes are displayed in solid red lines, the y-axes in dashed green lines and the z-axes in dash-dotted blue lines).}
\label{springmass} 
\end{figure}

As previously anticipated, docking mechanisms in reality do not act as simple 6 DOF clamped connections. For that purpose, local spring-dampers are used, implementing the model of a massless spring-damper system in order to close the kinematic chain of two rigid bodies at a particular point of the mechanism loop. The new connection model, which is displayed in Fig. \ref{springmass}a, replaces $\mathcal{C}_{1}$ and $\mathcal{C}_{2}$ by two spring-damper systems $\mathcal{SM}_{1}$ and $\mathcal{SM}_{2}$. It can also be observed that the dynamic model of the target's rigid hub is now given by ${[\textbf{I}_{D_{2}D_{3}P_{3}P_{4}}^{\mathcal{RH}_2}]}_{\mathcal{R}_{\mathcal{RH}_2}}$, since both spring-damper models output the wrenches to be applied to the rigid hub $\mathcal{RH}_{2}$. The dynamic model of a general spring-damper system $\mathcal{SM}_{\bullet}$, also displayed in {\Large \textcircled{\normalsize G}}, reads: 

\begin{equation}
\begin{aligned}
{\ddot{\mathbf{x}}_{C}}-{\ddot{\mathbf{x}}_{P}}={\delta{\ddot{\mathbf{x}}}}_{Q_{\bullet}} \quad \text{ and } \quad
\begin{cases}
  {\mathbf{W}_{\mathcal{SM}_{\bullet}/.,C}}=-\left({\textbf{K}_{{Q}_{\bullet}}} {\delta \mathbf{x}}_{Q_{\bullet}}+{\textbf{D}_{{Q}_{\bullet}}} {\delta \dot{\mathbf{x}}}_{Q_{\bullet}}\right)\\
  {\mathbf{W}_{\mathcal{SM}_{\bullet}/.,P}}={\textbf{K}_{{Q}_{\bullet}}} {\delta \mathbf{x}}_{Q_{\bullet}}+{\textbf{D}_{{Q}_{\bullet}}} {\delta \dot{\mathbf{x}}}_{Q_{\bullet}}
\end{cases}
\end{aligned}
\label{springmasseq}
\end{equation}

where ${\textbf{K}_{{Q}_{\bullet}}}=\left[\begin{array}{cc}
{\delta_{{K}_{\bullet}}^{shear}}\textbf{I}_{3} & 0_{3 \times 3} \\
0_{3 \times 3} & {\delta_{{K}_{\bullet}}^{tors}}\textbf{I}_{3}
\end{array}\right]$ and ${\textbf{D}_{{Q}_{\bullet}}}=\left[\begin{array}{cc}
{\delta_{{D}_{\bullet}}^{shear}}\textbf{I}_{3} & 0_{3 \times 3} \\
0_{3 \times 3} & {\delta_{{D}_{\bullet}}^{tors}}\textbf{I}_{3}
\end{array}\right]$ are uncertain stiffness and damping six-by-six matrices acting on the six different DOF, respectively. Furthermore, ${\delta_{{K}_{\bullet}}^{shear}}$ $[N/m]$ and ${\delta_{{D}_{\bullet}}^{shear}}$ $[Ns/m]$ are real parametric uncertainties describing the shear stiffness and damping properties of these spring-damper systems, respectively. Similarly, ${\delta_{{K}_{\bullet}}^{tors}}$ $[Nm/rad]$ and ${\delta_{{D}_{\bullet}}^{tors}}$ $[Nms/rad]$ are real parametric uncertainties describing the torsional stiffness and damping coefficients of the spring-dampers. Therefore, the uncertainty block of each spring-damper system is given by $\boldsymbol{\Delta}_{\mathcal{SM}_{\bullet}}=\operatorname{diag}\left({\delta_{{K}_{\bullet}}^{shear}}\textbf{I}_{3}, {\delta_{{K}_{\bullet}}^{tors}}\textbf{I}_{3}, {\delta_{{D}_{\bullet}}^{shear}}\textbf{I}_{3}, {\delta_{{D}_{\bullet}}^{tors}}\textbf{I}_{3}\right)$. Since ${\textbf{K}_{{Q}_{\bullet}}}$ and ${\textbf{D}_{{Q}_{\bullet}}}$ are considered uncertain, one can investigate the system's behaviour while docking/undocking occurs. Eq. (\ref{springmasseq}) is always written in the respective reference points $Q_{\bullet}$ of the systems $\mathcal{SM}_{\bullet}$. Furthermore, Eq. (\ref{springmasseq}) is also projected in reference frames $\mathcal{R}_{{\mathcal{SM}}_{\bullet}}=\left({Q_{\bullet}}; x_{Q_{\bullet}}, y_{Q_{\bullet}}, z_{Q_{\bullet}}\right)$, depicted in Fig. \ref{springmass}b.

The two docking phases of this OOS mission scenario are shown in Fig. \ref{springmass}b. Illustration {\Large \textcircled{\normalsize 7}} shows the moment where the robotic arm docks to the target spacecraft. Moreover, illustration {\Large \textcircled{\normalsize 8}} shows the instant where the target's rigid hub docks to the chaser's rigid hub, followed by the robotic arm disengaging from the target spacecraft. Two different LFR models are built in order to study each one of these two different docking phases, which are obtained by introducing the new connection system displayed in Fig. \ref{springmass}a in the block-diagram representation shown in Fig. \ref{complete_lft}.

\begin{figure}[!ht]
\centering
 \includegraphics[width=1\textwidth]{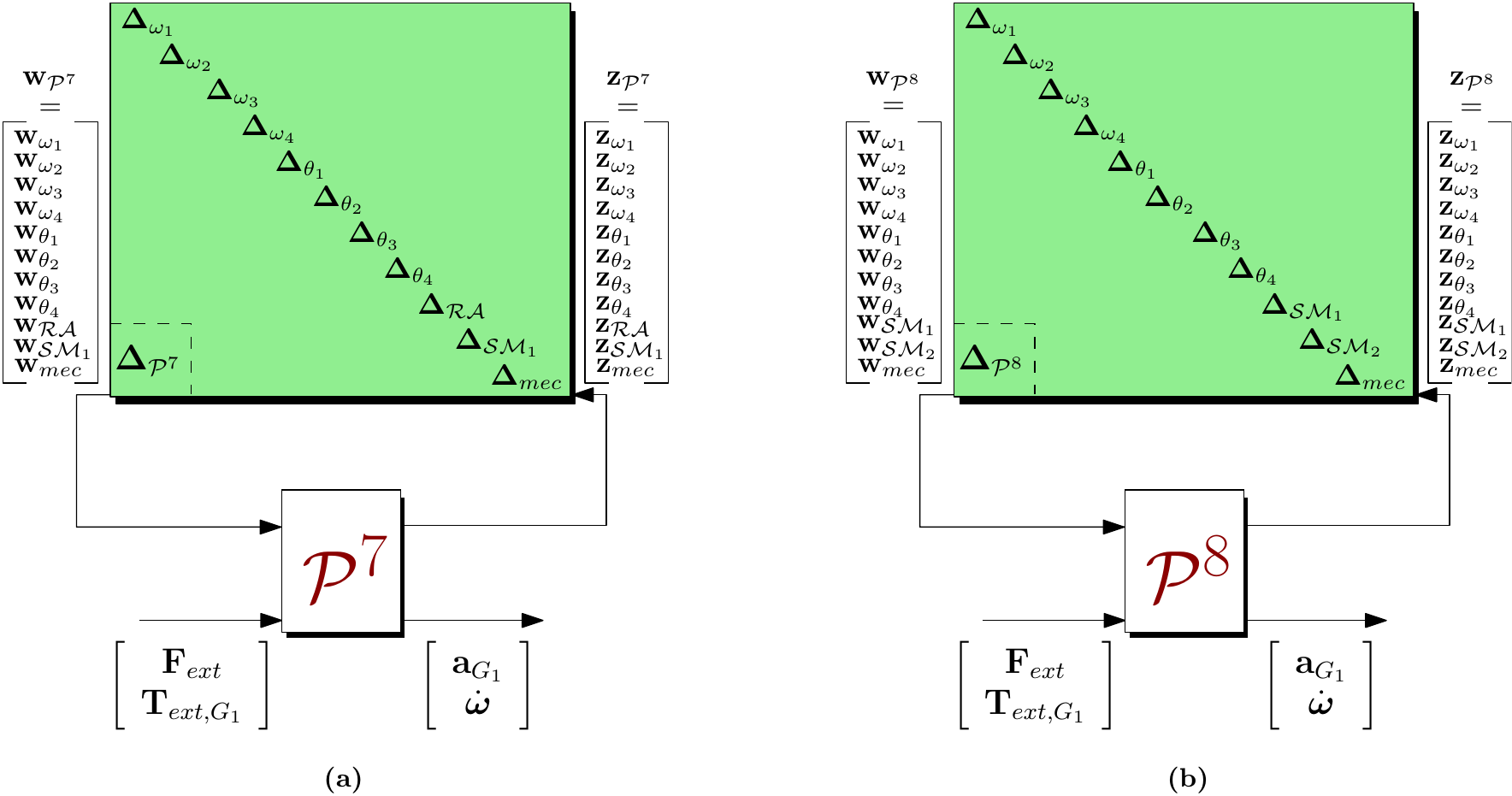}
\caption{Global LFR representations: (a) Docking phase {\normalsize \textcircled{\scriptsize 7}}. (b) Docking phase {\normalsize \textcircled{\scriptsize 8}}.} 
\label{compact_lft_springmass} 
\end{figure}

\textbf{Docking phase} {\Large \textcircled{\normalsize 7}}: Even though the robotic arm configuration is static and known in {\Large \textcircled{\normalsize 7}}, this model is also parameterized according to $\alpha_{\bullet}$, which gives the possibility to study this type of docking for different robotic arm configurations. However, this LFR model cannot be parameterized according to $\boldsymbol{\Delta}_{\mathcal{SM}_{2}}$, since the system is only in equilibrium when ${\textbf{K}_{{Q}_{2}}}=0_{6 \times 6}$ and  ${\textbf{D}_{{Q}_{2}}}=0_{6 \times 6}$, due to the fact that  $D_{1} \nequiv D_{3}$. A new global LFR form is thus obtained, as depicted in Fig. \ref{compact_lft_springmass}a.

\textbf{Docking phase} {\Large \textcircled{\normalsize 8}}: The connection model depicted in Fig. \ref{springmass}a is also used here. However, instant {\Large \textcircled{\normalsize 8}} differs from {\Large \textcircled{\normalsize 7}} in the sense that it introduces a closed-loop kinematic chain. Simply put, for closed-loop kinematic chain systems, the number of rigid and independent DOF is reduced due to the loop closure constraints:

\begin{itemize}

\item $\displaystyle \mathbf{g}\left(\alpha_{tot}, \mathbf{x}_{f}\right)=\mathbf0 \text { (on positions). }$

\item  $\displaystyle {\frac{\partial \mathbf{g}}{\partial {\alpha_{tot}}}}^{T} \dot{{\alpha_{tot}}}+{\frac{\partial \mathbf{g}}{\partial \mathbf{x}_{f}}}^{T} \dot{\mathbf{x}}_{f}=\mathbf0 \text { (on velocities). }$

\item $\displaystyle {\frac{\partial \mathbf{g}}{\partial {\alpha_{tot}}}}^{T} \ddot{{\alpha_{tot}}}+{\frac{\partial \mathbf{g}}{\partial \mathbf{x}_{f}}}^{T} \ddot{\mathbf{x}}_{f}+\underbrace{\frac{d}{d t}\left({\frac{\partial \mathbf{g}}{\partial {\alpha_{tot}}}}^{T}\right) \dot{{\alpha_{tot}}}+\frac{d}{d t}\left({\frac{\partial \mathbf{g}}{\partial \mathbf{x}_{f}}}^{T}\right) \dot{\mathbf{x}}_{f}}_{\text {non-linear terms }}=\mathbf0 \text { (on accelerations). }$
  
\end{itemize}

with ${\alpha_{tot}}=\left[\alpha_{1}, \alpha_{2}, \alpha_{3}, \alpha_{4}, \alpha_{5}, \alpha_{6}\right]^{T}$ and $\mathbf{x}_{f}$ being the deformation DOF vector of the whole system. These constraints $\mathbf{g}$ are non-linear equations which are not solved in SDT, which means that the system parameterized according to the robotic arm geometrical configuration cannot be derived in this case. It is however possible to provide a geometrical configuration of the robotic arm $\alpha_{ref}$ satisfying these constraints and the equilibrium conditions, i.e. $\mathbf{g}\left(\alpha_{ref}, \mathbf0\right)=\mathbf0$. Then, SDT can be used to compute a linear model which is valid for small variations around the geometrical configuration $\alpha_{ref}$ while satisfying the linearized loop closure constraints at the acceleration level:

\begin{equation}
{\frac{\partial \mathbf{g}}{\partial {\alpha_{tot}}}}^{T} \ddot{{\alpha_{tot}}}+{\frac{\partial \mathbf{g}}{\partial \mathbf{x}_{f}}}^{T} \ddot{\mathbf{x}}_{f}=\mathbf0
\end{equation}

In the circumstances displayed in {\Large \textcircled{\normalsize 8}}, the non-linear loop closure constraint at the acceleration level is given by ${[\ddot{\mathbf{x}}_{D_{1}}]}_{\mathcal{R}_{\mathcal{RH}_{1}}}= {[\tau_{D_{1}D_{3}}]}_{\mathcal{R}_{\mathcal{RH}_{1}}} {[\ddot{\mathbf{x}}_{D_{3}}]}_{\mathcal{R}_{\mathcal{RH}_{1}}}$, where ${[\tau_{D_{1}D_{3}}]}_{\mathcal{R}_{\mathcal{RH}_{1}}}$ depends on $\alpha_{tot}$. Nevertheless, for this on-orbit servicing mission, docking occurs when the robotic arm is static and $\alpha_{ref}$ is perfectly known. Since $D_{1} \equiv D_{3} \equiv Q_{2}$ and $J_{7} \equiv D_{2} \equiv Q_{1}$, the blocks $\boldsymbol{\Delta}_{\mathcal{SM}_{\bullet}}$ are left uncertain. A new global LFR form is derived, as depicted in Fig. \ref{compact_lft_springmass}b.

The 2 LFR models displayed in Fig. \ref{compact_lft_springmass} are thus obtained in order to model the system during the two existent docking phases that take place during the OOS mission scenario in question.

\subsection{Analysis of the system dynamics}

\begin{figure}[!ht]
\centering
 \includegraphics[width=1\textwidth]{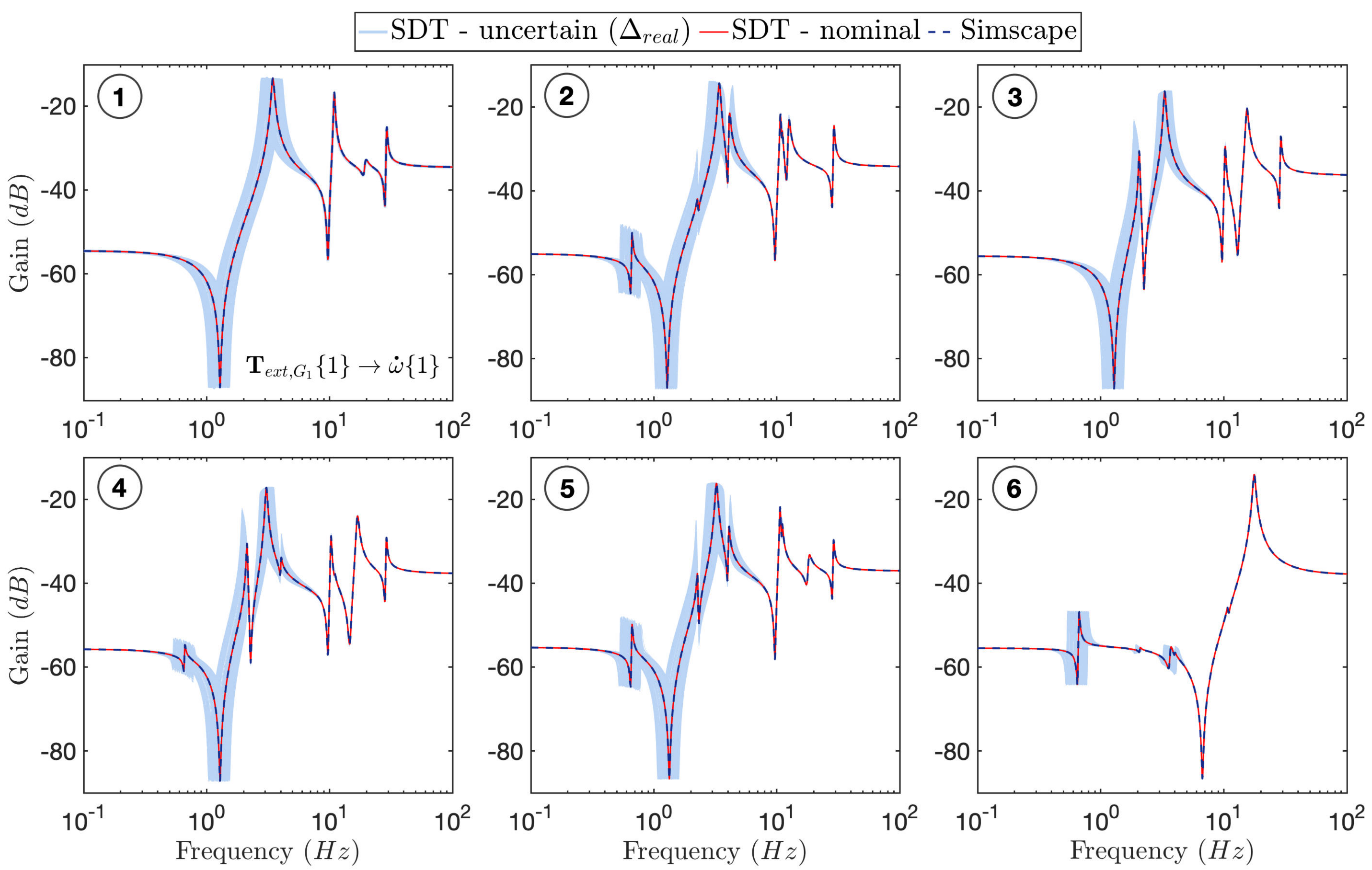}
\caption{Comparison between the gains of the SDT - uncertain, SDT - nominal and \textit{Simscape} systems for the six moments displayed in figure \ref{simscape_rep}; transfer between the first components of the external torque $\mathbf{T}_{e x t, G_{1}}$ and angular acceleration $\boldsymbol{\dot{{\omega}}}$ ($\mathbf{T}_{e x t, G_{1}}\{1\} \rightarrow \boldsymbol{\dot{{\omega}}}\{1\}$ channel).}
\label{sigma_plot} 
\end{figure}

Before proceeding to control design, it has to be ascertained whether the SDT open loop model represented in Fig. \ref{complete_lft} and the \textit{Simscape} system which was built in parallel are identical in the linear domain. Let us now compare the singular values between the \textit{Simscape} and nominal SDT systems for all six different configurations displayed in Fig. \ref{simscape_rep} considering the plant shown in Fig. \ref{compact_lft}. For instance, if the transfer function from the first component of the external torque $\mathbf{T}_{e x t, G_{1}}$ to the first component of the angular acceleration $\boldsymbol{\dot{{\omega}}}$ is considered, Fig. \ref{sigma_plot} shows an excellent match, since the red and dashed blue lines overlap. The plot is also coherent with the properties of both spacecraft's flexible elements, in accordance with the modal participation factor matrices $\mathbf{L}_{P_{\bullet}}^{\mathcal{SA}_{\bullet}}$ definition \cite{Guy2014}. Indeed, the antiresonances occur at the frequencies of the cantilevered flexible modes corresponding to the solar arrays. Since SDT computes the inverse linearized dynamic model of the whole system projected in $\mathcal{R}_{\mathcal{RH}_{1}}$ and the channel  $\mathbf{T}_{e x t, G_{1}}\{1\} \rightarrow \boldsymbol{\dot{{\omega}}}\{1\}$ is being analyzed, the static gains of all the plots in Fig. \ref{sigma_plot} represent the inverse of the first moment of inertia $1/\textbf{J}_{xx}$ measured at point $G_{1}$ and with respect to $\mathcal{R}_{\mathcal{RH}_{1}}$ for all the different moments depicted in Fig. \ref{simscape_rep}.

Furthermore, Fig. \ref{sigma_plot} also depicts the effect that the set of real parametric uncertainties $\boldsymbol{\Delta}_{real}=\operatorname{diag}\left(\boldsymbol{\Delta}_{mec}, \boldsymbol{\Delta}_{mod}\right)$ has on the singular values of the channel $\mathbf{T}_{e x t, G_{1}}\{1\} \rightarrow \boldsymbol{\dot{{\omega}}}\{1\}$. Logically, all the varying tilt angles $\theta_{\bullet}$ and $\alpha_{\bullet}$ are set according to the geometrical configurations that can be observed in the illustrations of Fig. \ref{simscape_rep}. The parametric uncertainties $\delta_{\mathcal{C}_{\bullet}}$ are also replaced accordingly. 

\begin{figure}[!ht]
\centering
 \includegraphics[width=1\textwidth]{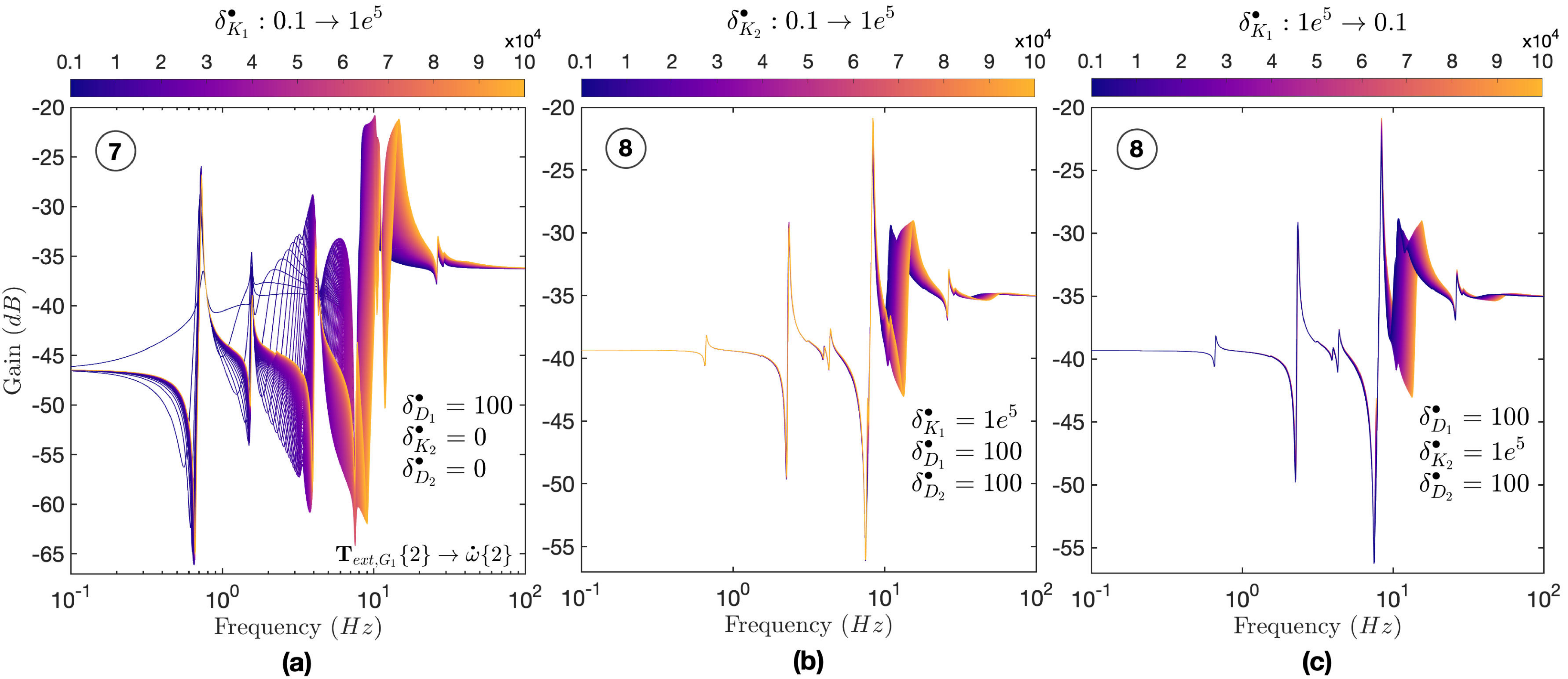}
\caption{Gains of the channel $\mathbf{T}_{e x t, G}\{2\} \rightarrow \boldsymbol{\dot{{\omega}}}\{2\}$ for moments  {\normalsize \textcircled{\scriptsize 7}} and  {\normalsize \textcircled{\scriptsize 8}} when considering two spring-damper systems: (a) the robotic arm docks to the target spacecraft. (b) the target docks to the chaser. (c) the robotic arm undocks from the target (Note: The units of $\delta_{{K}_{\bullet}}^{\bullet}$ are $[N/m]$ or $[Nm/rad]$ and the units of $\delta_{{C}_{\bullet}}^{\bullet}$ are $[Ns/m]$ or $[Nms/rad]$, depending on whether the stiffness and damping coefficients are shear or torsional).}
\label{sigma_plot_springmass} 
\end{figure}

Moreover, Fig. \ref{sigma_plot_springmass} assesses the behaviour of the channel $\mathbf{T}_{e x t, G_{1}}\{2\} \rightarrow \boldsymbol{\dot{{\omega}}}\{2\}$ when docking happens. For that reason, the LFR models depicted in Fig. \ref{compact_lft_springmass} are used. The tilt angles $\theta_{\bullet}$ and $\alpha_{\bullet}$ are replaced accordingly and all the modal and mechanical uncertainties $\boldsymbol{\Delta}_{real}$ are set as nominal. When the damping and stiffness coefficients are very big, a spring-damper system behaves just like a clamped attachment. In addition, when a body docks to another one, the stiffness coefficients are initially small and increase with respect to time.

Fig. \ref{sigma_plot_springmass}a shows how the system behaves when the robotic arm docks to the target spacecraft, where the shear and torsional damping coefficients of $\mathcal{SM}_{1}$ are equal to $\delta_{{D}_{1}}^{\bullet}=100$ $([Ns/m]$ or $[Nms/rad])$ and the stiffness coefficients $\delta_{{K}_{1}}^{\bullet}$ vary from $0.1$ to $1e^5$ $([N/m]$ or $[Nm/rad])$. Since $\mathcal{SM}_{2}$ is disconnected, $\delta_{{D}_{2}}^{\bullet}=0$ $([Ns/m]$ or $[Nms/rad])$  and $\delta_{{K}_{2}}^{\bullet}=0$ $([N/m]$ or $[Nm/rad])$. When $\delta_{{K}_{1}}^{\bullet}$ are small, the flexible modes of the target's solar arrays show very small or even non-existent antiresonances/resonances. This can be explained by the fact that the connection to the target is still very weak when $\delta_{{K}_{1}}^{\bullet}$ are small and therefore the flexible modes do not have a relevant effect on the system dynamics. As the coefficients $\delta_{{K}_{1}}^{\bullet}$ increase, these resonances and antiresonances start showing up. One of the examples is the antiresonance at around 0.65 $Hz$, which corresponds to the first flexible mode of the target's solar arrays. Since $\delta_{{D}_{1}}^{\bullet}=100$ $([Ns/m]$ or $[Nms/rad])$, the six flexible modes which are introduced by $\mathcal{SM}_{1}$ are noticeable even when $\delta_{{K}_{1}}^{\bullet}$ are very big. Logically, as the coefficients $\delta_{{K}_{1}}^{\bullet}$ increase, there is a shift of these flexible modes to the right, as can be observed in Fig. \ref{sigma_plot_springmass}a. 

Moreover, Fig. \ref{sigma_plot_springmass}b and \ref{sigma_plot_springmass}c show both phases of the second docking. First, $\delta_{{K}_{2}}^{\bullet}$ increase in order to mimic the docking of the target spacecraft to the rigid hub. Then, once this connection is rigid, $\delta_{{K}_{1}}^{\bullet}$ decrease in order to detach the robotic arm from the target's rigid hub. The damping coefficients of $\mathcal{SM}_{2}$ are equal to $\delta_{{D}_{2}}^{\bullet}=100$ $([Ns/m]$ or $[Nms/rad])$. In this case, the flexible modes which are introduced by $\mathcal{SM}_{\bullet}$ do not interfere with the system dynamics for frequencies smaller than 10 $Hz$. In addition, it can also be observed that the evolution of the system is similar for both cases when $\delta_{{K}_{2}}^{\bullet}$ increase and when $\delta_{{K}_{1}}^{\bullet}$ decrease. Another aspect which can be perceived is that the static gain of the singular values plot shown in Fig. \ref{sigma_plot_springmass}a is smaller than the one shown in Fig. \ref{sigma_plot_springmass}b and \ref{sigma_plot_springmass}c. This makes sense since the singular values correspond to the channel $\mathbf{T}_{e x t, G}\{2\} \rightarrow \boldsymbol{\dot{{\omega}}}\{2\}$ and the second moment of inertia of the coupled system $\textbf{J}_{yy}$ measured at point $G_{1}$ with respect to $\mathcal{R}_{\mathcal{RH}_{1}}$ is bigger in {\Large \textcircled{\normalsize 7}} than in {\Large \textcircled{\normalsize 8}}.

\subsection{Time-varying analysis}

\begin{figure}[!ht]
\centering
 \includegraphics[width=1\textwidth]{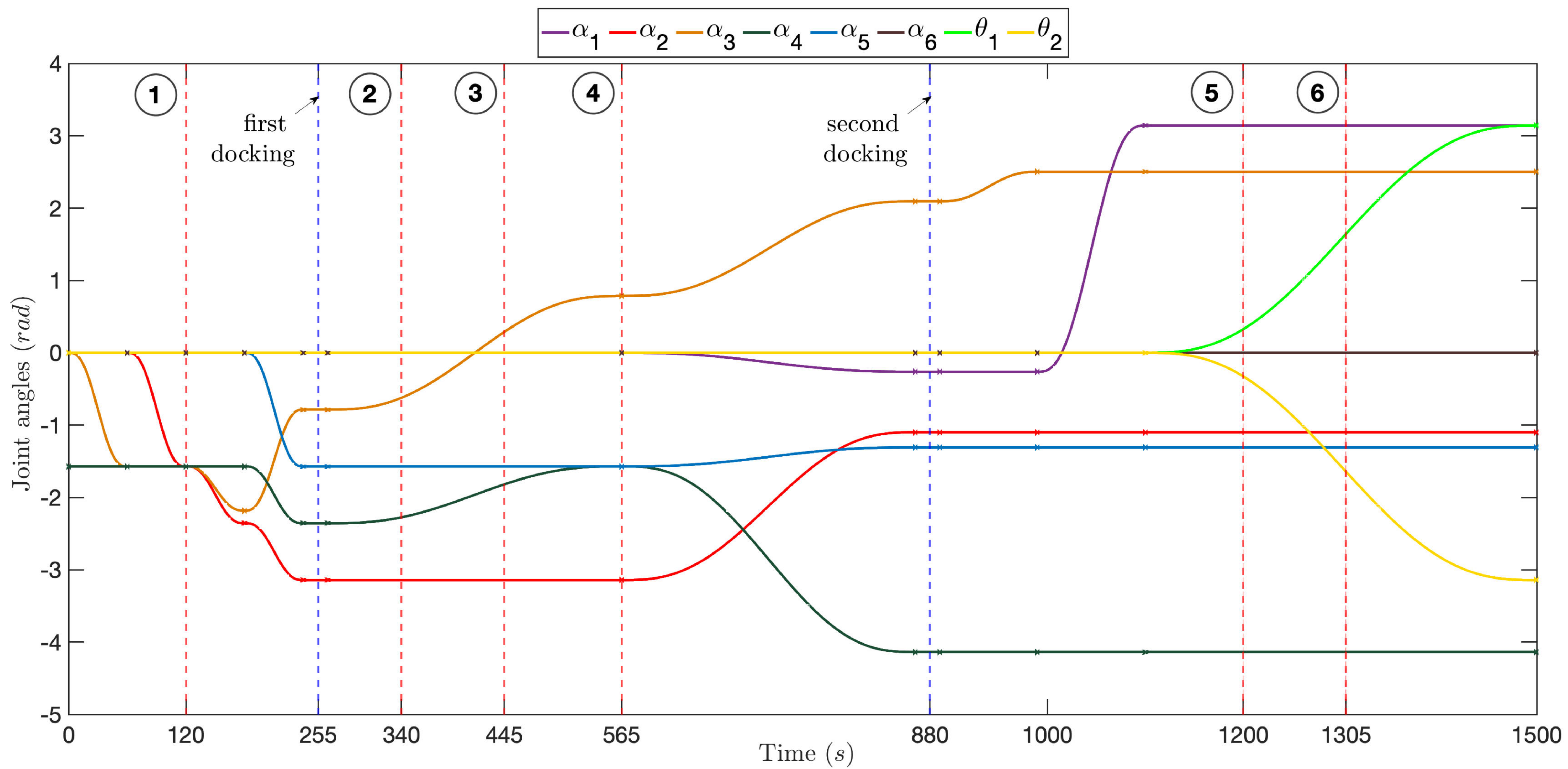}
\caption{Geometrical configuration of the robotic arm $\alpha_{\bullet}$ and chaser's solar arrays $\theta_{1,2}$ computed along 1500 seconds.}
\label{arm_traj} 
\end{figure}

By means of both connection models, this system fully captures the dynamics and interactions between all subsystems as well as the decoupled/coupled configurations. Nevertheless, the trajectories of the robotic arm and solar arrays still need to be defined. With that purpose, fifth-order polynomials are generated, which achieve a given set of input waypoints expressed in terms of joint configurations. The result is displayed in Fig. \ref{arm_traj}. It should also be noted that the target's solar panels are considered to be static. Henceforth, $\theta_{3}$ and $\theta_{4}$ are not represented in Fig. \ref{arm_traj} and are both equal to $0$ $rad$. 

\begin{figure}[!ht]
\centering
 \includegraphics[width=1\textwidth]{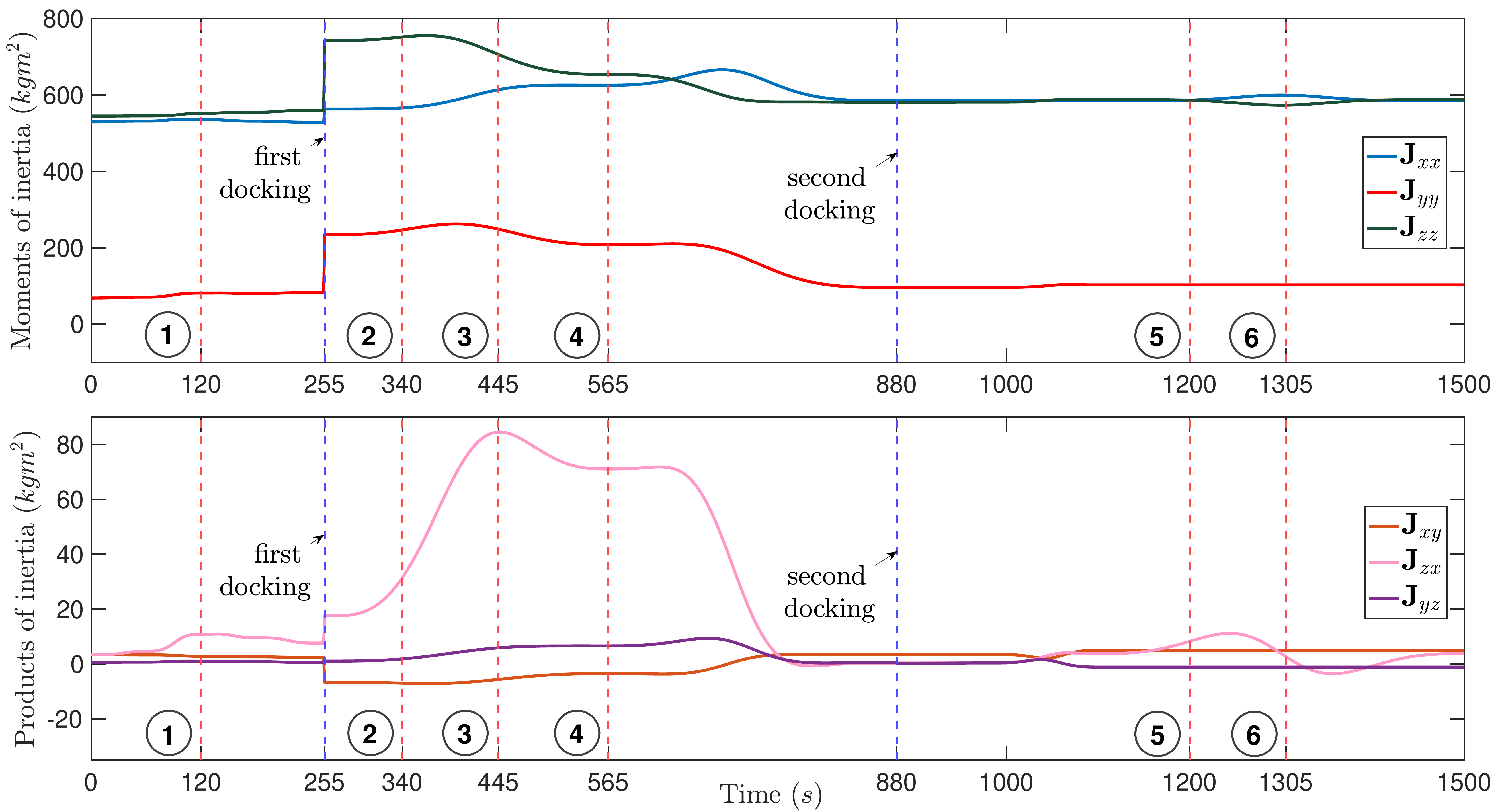}
\caption{Evolution of the nominal open loop system's inertia tensor entries with respect to time, computed at $G_{1}$ with regard to frame $\mathcal{R}_{\mathcal{RH}_{1}}$.}
\label{inertias} 
\end{figure}

The first thing to notice is that the robotic arm starts unfolding during the approach phase, before the first docking occurs at around $t=255$ $s$. Afterwards, the target is attached to the chaser's bottom surface at $t=880$ $s$. Next, the robotic arm undocks from the target and moves to another configuration. Finally, the chaser's solar arrays start tilting at $t=1100$ $s$. For better analyzing the behaviour of the nominal open loop system, the evolution of the products and moments of inertia with respect to time is obtained and depicted in Fig. \ref{inertias}. It should also be noted that these inertias are measured with respect to  $\mathcal{R}_{\mathcal{RH}_{1}}$. In fact, this plot outlines why it is so fundamental to parameterize the system with respect to its geometrical configuration and to be able to take these inertial changes into account when applying a control methodology. When the first docking occurs, an increase in the moments and products of inertia can be easily observed. However, as the robotic arm moves and brings the target closer to the chaser's rigid body, the inertia tensor entries start approaching their original values. Around moments {\Large \textcircled{\normalsize 5}} and {\Large \textcircled{\normalsize 6}}, some variations can be noticed due to the movement of the solar arrays. Since these rotations happen around $y_{P_{1}}$ and $y_{P_{2}}$, $\textbf{J}_{yy}$ stays constant.

\begin{figure}[!ht]
\centering
 \includegraphics[width=1\textwidth]{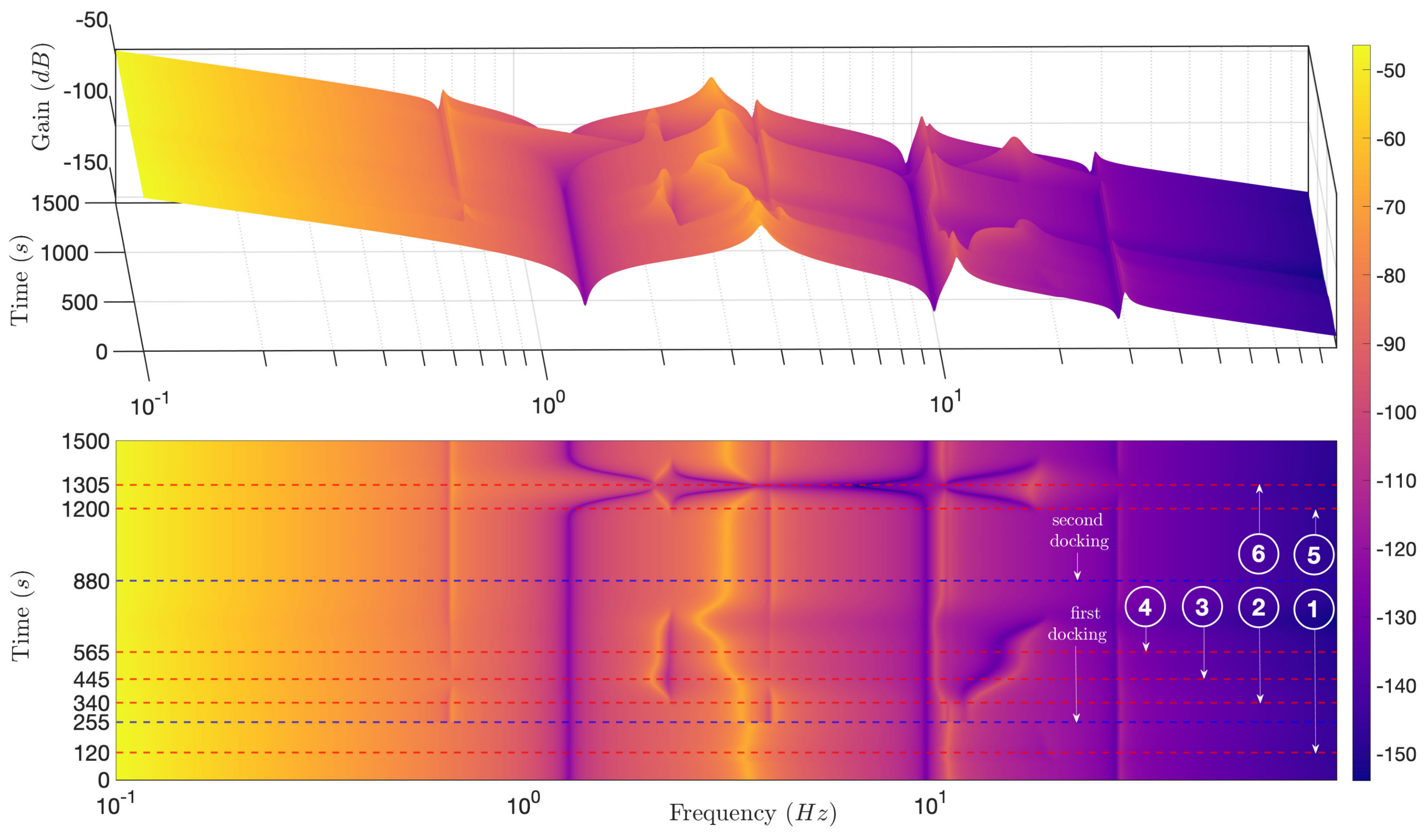}
\caption{Singular values of the nominal open loop system  with respect to time and along a dense grid of frequencies; transfer between the first components of the external torque $\mathbf{T}_{e x t, G_{1}}$ and linearized euler angles of the main body $\boldsymbol{\phi}$ ($\mathbf{T}_{e x t, G_{1}}\{1\} \rightarrow \boldsymbol{\phi}\{1\}$ channel).}
\label{surf} 
\end{figure}

Moreover, by inspecting the singular values for different frequencies and system configurations, Fig. \ref{surf} shows in a clear manner how the system's flexible modes evolve when the robotic arm moves and the tilt angles of the solar arrays vary. The channel $\mathbf{T}_{e x t, G_{1}}\{1\} \rightarrow \boldsymbol{\phi}\{1\}$ is considered, where $\boldsymbol{\phi}=\left[\begin{array}{lll}\phi_{x} & \phi_{y} & \phi_{z}\end{array}\right]^{T}$ denotes the linearized euler angles of the main body $\mathcal{RH}_{1}$ with respect to the inertial frame $\mathcal{R}_{\mathcal{O}}$. It should also be noted that $\boldsymbol{\phi}$ is obtained from a double integration of $\boldsymbol{\dot{\omega}}$.

Before the first docking, the flexible modes remain practically constant, since the movement of the robotic arm is almost negligible. However, after the first docking occurs at $t=255$ $s$, some of the flexible modes in the mid-range frequency show a very interesting evolution, which can be explained by the fact that the robotic arm is bringing the target spacecraft closer to the chaser's rigid hub, thus causing great changes in the inertial characteristics of the coupled system. Next, a stabilization can be observed after $t=880$ $s$, since the second docking takes place and the robotic arm movement does not provoke big changes in the mechanical parameters. Finally, when the chaser's solar arrays start tilting at $t=1100$ $s$, a symmetric evolution of some flexible modes can be observed. This behaviour can be explained by the fact that the chaser's solar arrays are almost symmetric and they undergo a revolution of 180 degrees. 

\section{Control architecture and synthesis methodology}
\label{control}

\subsection{Baseline attitude controller}

Some of the challenges of an OOS mission scenario include the control structure interactions between the flexible appendages and the AOCS, the time-varying inertial properties, the flexible dynamics, the system uncertainties and also the dynamic couplings. Initially, the control design is divided into two different parts, namely the translational and the attitude. However, only attitude control design is addressed, since this paper's objective is to focus on the very last phase of the rendezvous between the chaser and target spacecraft. A baseline attitude controller is tuned based on the inertial properties of the coupled system when the first docking takes place at $t=255$ $s$, as follows: 

\begin{equation}
\mathbf{u}=\boldsymbol{K}_{att}\left[\begin{array}{c}
\boldsymbol{\phi}_{ref} - \boldsymbol{\phi}\\
\boldsymbol{\omega}_{ref} - \boldsymbol{\omega}
\end{array}\right] \quad \text{and} \quad
\boldsymbol{K}_{att}=\left[\begin{array}{ll}\boldsymbol{k}_{att} & \boldsymbol{c}_{att}\end{array}\right]
\quad \text{with} \quad
\begin{cases}
  \boldsymbol{k}_{att}= \omega_{att}^{2} \textbf{J}_{tot}\\
  \boldsymbol{c}_{att}= 2 \xi_{att} \omega_{att} \textbf{J}_{tot}
\end{cases}  
\end{equation}

where $\mathbf{u}$ represents the torque control output and $\boldsymbol{\omega}$ is the angular velocity of the chaser spacecraft defined in $\mathcal{R}_{\mathcal{RH}_{1}}$ with respect to the inertial frame $\mathcal{R}_{O}=\left(O; x_{O}, y_{O}, z_{O}\right)$ and obtained from integrating $\boldsymbol{\dot{\omega}}$. Moreover, $\textbf{J}_{tot}$ is the inertia tensor of the collection of all the body elements, measured with respect to $\mathcal{R}_{\mathcal{RH}_{1}}$. The objective is to have a critically damped system that returns to rest slowly without oscillating when tracking the reference signals $\boldsymbol{\phi}_{ref}$ and $\boldsymbol{\omega}_{ref}$. For that reason, $\xi_{att}=1$ and $\omega_{att}=0.01$ $Hz$ represent the controller's damping ratio and natural frequency, respectively. 

The attitude controller is to take action during the complete mission scenario, which involves big changes in the orientation and magnitude of the inertia tensor. Consequently, this tuning methodology is far from appropriate when considering such a challenging and complex scenario. Nevertheless, this controller is useful as an initial guess when using more powerful and advanced control methodologies, such as $\mathcal{H}_{\infty}$, which is able to optimize a first approximation until reaching the limits of robustness and performance.

\subsection{$\mathcal{H}_{\infty}$ control}

\begin{figure}[!ht]
\centering
 \includegraphics[width=1\textwidth]{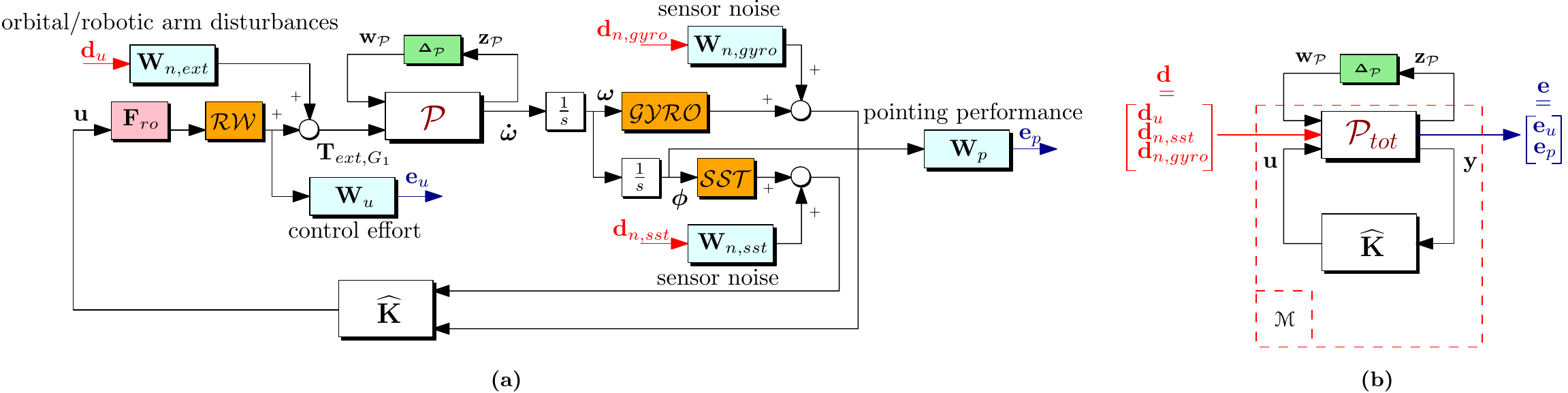}
\caption{(a) System architecture used for controller synthesis and worst-case analysis. (b) Equivalent standard form of the interconnection.}
\label{controldesign} 
\end{figure}

In order to design a control law that accommodates the desired performance requirements, the synthesis problem is recast into the nonsmooth $\mathcal{H}_{\infty}$ framework \cite{Apkarian2006} by first assembling the weighted interconnection shown in Fig. \ref{controldesign}a. First, the plant model $\mathcal{P}$ represented in Fig. \ref{compact_lft} is introduced. However, only the $\mathbf{T}_{e x t, G_{1}} \rightarrow \boldsymbol{\dot{\omega}}$ channels are considered, since the objective is to improve the attitude controller. This interconnection is composed of the following blocks:

\begin{itemize}

\item \textbf{Sensor and actuator models}: First, the star tracker dynamics $\mathcal{SST}$ corresponds to a first order low pass filter with a cutoff frequency of 8 $Hz$. Secondly, the gyroscope dynamics $\mathcal{GYRO}$ is represented by a first order low pass filter with a 200 $Hz$ cutoff frequency. Finally, the reaction wheel system dynamics $\mathcal{RW}$ is approximated by a second order transfer, with a damping ratio equal to 0.7 and a natural frequency of 200 $Hz$.

\item \textbf{Disturbance weights $\mathbf{W}_{n,gyro}$, $\mathbf{W}_{n,sst}$} and $\mathbf{W}_{n,ext}$: The measurement noise weights $\mathbf{W}_{n,gyro}$ and $\mathbf{W}_{n,sst}$ are used to define the upper bounds on the expected spectral amplitude of the closed-loop noise measurements. In this case, $\mathbf{W}_{n,gyro} =  9.1987e^{-04}\textbf{I}_{3}$ $rad/s$  and $\mathbf{W}_{n,sst} =  1.5343e^{-05}\textbf{I}_{3}$ $rad$. Similarly, the purpose of the weight $\mathbf{W}_{n,ext}$ is to model the upper bound on the expected closed-loop orbital and robotic arm disturbances at different frequencies. Even though the control loop should include a feedforward term responsible for counteracting the resistance against motion on the locked axes of the joint connecting the robotic arm and the chaser's rigid hub, some residual constraint torques will still continue to exist. Therefore, the disturbance weight $\mathbf{W}_{n,ext}=\operatorname{diag}\left(\frac{0.002577}{2.236s+0.2236},\, \frac{0.009685}{2.236s+0.2236},\, \frac{0.01239}{2.236s+0.2236}\right)$ $Nm$ is tuned to take into account these robotic arm disturbances and also orbital perturbations acting on the system, like magnetic or gravity gradient torques.

\item \textbf{Performance weights $\mathbf{W}_{u}$ and $\mathbf{W}_{p}$}: The purpose of the weight $\mathbf{W}_{u}=\operatorname{diag}\left(0.5,\, 0.5,\, 0.5\right)$ $Nm$ is to impose a desired closed-loop upper bound of 2 $Nm$ on the worst-case actuator signals at different frequencies. Moreover, the Absolute Pointing Error (APE) requirement is given by $\mathbf{W}_{p} = \operatorname{diag}\left(0.0079,\, 0.0079,\, 0.0079\right)$ $rad$.

\item \textbf{Roll-off filter} $\mathbf{F}_{ro}$: A 4th-order roll-off Butterworth filter with a cutoff frequency of 0.7 $Hz$ is also added to the output control signal $\mathbf{u}$, ensuring the controller is not sensitive to high frequency content.

\item \textbf{Structured controller $\widehat{\mathbf{K}}$}: The desired ACS (Attitude Control System) is a static and structured three-by-six controller. All 18 tunable gains are initially set to the values obtained with the baseline controller, which tuning is based on the initial inertial properties of the coupled system. The closed-loop model, denoted $\mathscr M$, is achieved with the lower linear fractional transformation $\mathcal{F}_{l}(\cdot)$ \cite{Preda2020} between the open loop model and the tunable three-by-six controller. This interconnection can be observed in Fig. \ref{controldesign}b, where $\mathscr M=\mathcal{F}_{l}\left(\mathcal{P}_{tot}, \widehat{\mathbf{K}}\right)$. Ultimately, the uncertain closed-loop model is given by $\mathscr M_{unc}=\mathcal{F}_{u}\left(\mathscr M, \boldsymbol{\Delta}_{\mathcal{P}}\right)$, where $\mathcal{F}_{u}(\cdot)$ represents the upper linear fractional transformation.

\end{itemize}{}

First, an array of 200 different plant models ${\mathscr M}_{200}$ is obtained from $\mathscr M_{unc}$, by replacing $\theta_{\bullet}$ and $\alpha_{\bullet}$ with 200 equally distributed geometrical configuration waypoints in the time domain. These plant models are obtained with the connection model depicted in Fig. \ref{compact_lft}, where ${\delta}_{\mathcal{C}_{1}}$ and ${\delta}_{\mathcal{C}_{2}}$ are also substituted accordingly. A soft constraint is considered on the $\mathcal{H}_{\infty}$ norm between the normalized disturbances $\mathbf{d}$ and outputs $\mathbf{e}$ of the system, as can be seen below:

\begin{equation}
\min _{\widehat{\mathbf{K}}}\gamma \quad \text{so that} \quad
\max _{\boldsymbol{\Delta_{real}}}\left\|{\mathscr M}_{200}\right\|_{\infty} \leq \gamma
\label{controlreq}
\end{equation}

Following Eq. (\ref{controlreq}), any controller for which $\gamma < 1$ satisfies the robust performance and stability requirements. The control methodology is finally applied on the array of 200 different plant models, where $\boldsymbol{\Delta}_{real}$ remains uncertain. In the end, this multimodel control design approach pretends to optimize a controller while taking into account all the big inertia changes that happen during the OOS mission scenario that is being studied. In the case of the optimization shown in Eq. (\ref{controlreq}), a controller $\widehat{\mathbf{K}}$ was found to achieve a performance level of $\gamma = 0.4$, meaning that the soft constraint has been completely satisfied. This control synthesis methodology allows for the design of a structured static controller in one shot, which is able to comply with the imposed requirements for all the different plant models.  An in-depth worst-case analysis is now provided in section \ref{analysis}.

\section{Performance and stability analysis}
\label{analysis}
\subsection{Worst-case analysis}

Let us now introduce the structured singular value function $\mu_{\delta}(\cdot)$ \cite{Doyle1993}, which provides very precise information about the magnitude of uncertainty which is needed to destabilize the loop at any frequency \cite{Preda2020}. If the nominal system (i.e. the block $\mathscr{M}_{\mathbf{d} \rightarrow \mathbf{e}}$) shown in Figure \ref{controldesign}b is stable, then the stability of this loop is conditioned by the existence of $\left(\textbf{I}-\mathscr{M}_{\mathbf{w}_{p} \rightarrow \mathbf{z}_{p}} \boldsymbol{\Delta}\right)^{-1}$. This is assessed by evaluating for different frequencies $\omega_{\mu} \in \mathbb{R}$ the structured singular value $\mu_{\delta}\left(\mathscr{M}_{\mathbf{w}_{p} \rightarrow \mathbf{z}_{p}}(j \omega_{\mu})\right)$, with $\mu_{\delta}(\cdot)$ being defined for a complex matrix $\mathscr{M} \in \mathbb{C}^{n \times m}$ and a set of uncertainties $\boldsymbol{\Delta} \in \Delta \subset \mathbb{R} \mathbb{H}_{\infty}^{m \times n}$ as:

\begin{equation}
\mu_{\delta}(\mathscr{M})=\frac{1}{\min \{\bar{\sigma}(\boldsymbol{\Delta}): \boldsymbol{\Delta} \in {\Delta}, \operatorname{det}(\textbf{I}-\mathscr{M} \boldsymbol{\Delta})=0\}}
\label{mu}
\end{equation}

where $\mathbb{C}^{n \times m}$ is the set of $n$-by-$m$ complex matrices, $\bar{\sigma}\left({\boldsymbol{\Delta}}\right)$ represents the maximum singular value of $\boldsymbol{{\Delta}}$ and the set $\mathbb{R} \mathbb{H}_{\infty}^{m \times n}$ describes the set of finite gain transfer matrices with $m$ outputs and $n$ inputs. For $\mathcal{G} \in \mathbb{R} \mathbb{H}_{\infty}^{m \times n}$, the value $\|\mathcal{G}\|_{\infty}$ represents the $\mathscr{L}_{2}$ system gain. If no $\boldsymbol{\Delta} \in {\Delta}$ makes $\textbf{I}-\mathscr{M} \boldsymbol{\Delta}$ singular, then $\mu_{\delta}(\mathscr{M}):=0$. Following this definition and under the assumption that the nominal system $\mathscr{M}_{\mathbf{d} \rightarrow \mathbf{e}}$ is stable, then $\mathcal{F}_{u}(\mathscr{M}, \boldsymbol{\Delta})$ is stable $\forall \boldsymbol{\Delta} \in {\Delta}$, $\bar{\sigma}(\boldsymbol{\Delta})<\nu$ if and only if: $\mu_{\delta}\left(\mathscr{M}_{\mathbf{w}_{p} \rightarrow \mathbf{z}_{p}}(j \omega_{\mu})\right)<1 / \nu ; \forall \omega_{\mu} \in \mathbb{R}$. In this context, $\mu_{\delta}$ gives a measure of the smallest structured uncertainty $\boldsymbol{\Delta}$ that causes closed-loop instability for any frequency $\omega_{\mu} \in$ $\mathbb{R}$. Moreover, the $\mathscr{L}_{2}$ gain of this destabilizing perturbation is exactly $1 / \mu_{\delta}$. This fact will be used to evaluate the stability margin of the loop with respect to different uncertainty structures in the analysis phase. However, due to its non-convex character, $\mu_{\delta}$ can be difficult to compute exactly. For that reason, some very efficient algorithms \cite{balas2007robust} have been developed in order to estimate the bounds of $\mu_{\delta}$. In order to assess the stability and performance robustness of the system, two further uncertainties are also considered:

\begin{itemize}

\item \textbf{Additive uncertainty $\boldsymbol{\Delta}_{add}$}: The additive uncertainty matrix $\boldsymbol{\Delta}_{add} \subset \mathbb{R} \mathbb{H}_{\infty}^{3 \times 3}$ is an unstructured uncertainty block with a very clear physical interpretation, since it maps the torque signals $\mathbf{T}_{e x t, G_{1}}$ to the angular acceleration signals $\boldsymbol{\dot{{\omega}}}$. Consequently, $\boldsymbol{\Delta}_{add}$ essentially covers for unknown inverse dynamic (static or flexible) inertia along the three axes. This 3-by-3 matrix is full (non-zero off-diagonal terms), therefore accounting for possible unknown cross-couplings between the different axes. These imprecisions are covered by the following model :

\begin{equation}
\hat{\mathcal{P}}=\mathcal{P}+\mathbf{W}_{a d d} \boldsymbol{\Delta}_{add} \quad \text { with } \quad  \mathbf{W}_{a d d}= -75 \textbf{I}_3 \, dB 
\end{equation}

The weight $\mathbf{W}_{a d d}$ is used to scale the magnitude of the additive normalized LTI uncertainty block $\boldsymbol{\Delta}_{add}$, with $\left\|\boldsymbol{\Delta}_{a d d}\right\|_{\infty} \leq 1$.

\item \textbf{Multiplicative uncertainty $\boldsymbol{\Delta}_{mul}$}: The multiplicative uncertainty block $\boldsymbol{\Delta}_{mul}=\operatorname{diag}\left(\boldsymbol{\Delta}_{mul_{L}},\boldsymbol{\Delta}_{mul_{R}}\right)$ is used to model neglected dynamics, gain fluctuations in the actuators and also phase uncertainty. The new uncertain control torques to the plant ${\mathbf{\hat{T}}_{e x t, G_{1}}}$ are equal to:

\begin{equation}
{\mathbf{\hat{T}}_{e x t, G_{1}}}=\left(\textbf{I}_{3}+\mathbf{W}_{{mul_{L}}}\boldsymbol{\Delta}_{mul_{L}}+\mathbf{W}_{{mul_{R}}}\boldsymbol{\Delta}_{mul_{R}}\right) \mathbf{T}_{e x t, G_{1}} \quad \text { with } \quad
\begin{array}{c}
\boldsymbol{\Delta}_{mul_{L}}=\left[\begin{array}{ccc}\delta_{mul_{xx}} & 0 & 0 \\ 0 & \delta_{mul_{yy}} & 0 \\ 0 & 0 & \delta_{mul_{zz}}\end{array}\right] \\ \\
\boldsymbol{\Delta}_{mul_{R}}=\left[\begin{array}{ccc}0 & \delta_{mul_{xy}} & \delta_{mul_{xz}}\\ \delta_{mul_{yx}} & 0 & \delta_{mul_{yz}} \\ \delta_{mul_{zx}} & \delta_{mul_{zy}} & 0 \end{array}\right]
\end{array}
\end{equation}

where $\delta_{mul_{\bullet}}$ are scalar normalized LTI uncertainties satisfying $\bar{\sigma}\left(\delta_{mul_{\bullet}}\right) \leq 1$. In addition, the weights $\mathbf{W}_{m u l_{L}}=\operatorname{diag}\left(4e^{-2},\, 4e^{-2},\, 4e^{-2}\right)$ and $\mathbf{W}_{m u l_{R}}=\operatorname{diag}\left(4e^{-3},\, 4e^{-3},\, 4e^{-3}\right)$ are used to scale the magnitude of the multiplicative LTI uncertainty blocks $\boldsymbol{\Delta}_{mul_{L}}$ and $\boldsymbol{\Delta}_{mul_{R}}$. In addition, it should also be noted that the diagonal terms of the expression $\mathbf{W}_{{mul_{L}}}\boldsymbol{\Delta}_{mul_{L}}+\mathbf{W}_{{mul_{R}}}\boldsymbol{\Delta}_{mul_{R}}$ are one order of magnitude larger than the off-diagonal terms, with the objective of considering possible couplings due to unmodeled effects or actuator misalignments.

\end{itemize}

\begin{figure}[!ht]
\centering
 \includegraphics[width=1\textwidth]{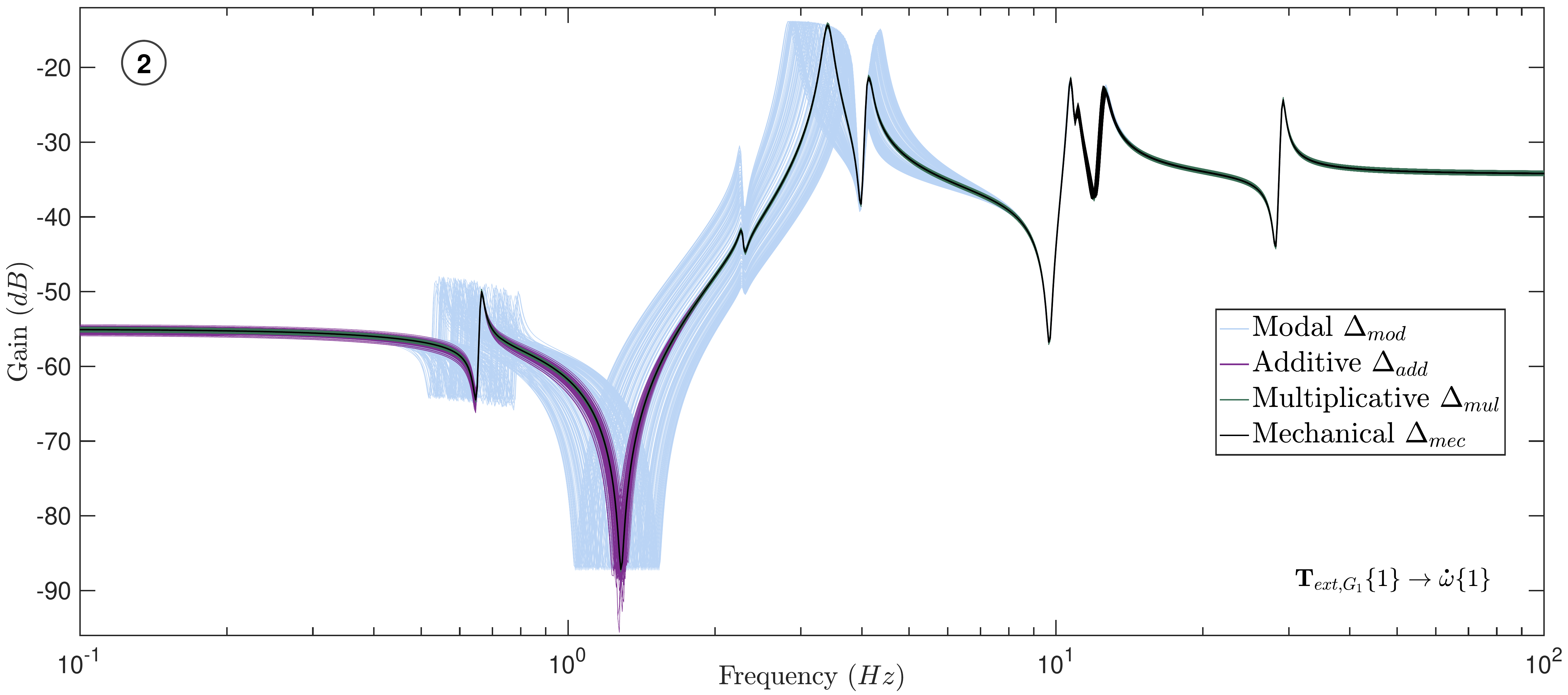}
\caption{The effects of different sets of uncertainties on the gains of the channel $\mathbf{T}_{e x t, G_{1}}\{1\} \rightarrow \boldsymbol{\dot{{\omega}}}\{1\}$ for instant {\normalsize \textcircled{\scriptsize 2}}.}
\label{unc_sensitivity} 
\end{figure}

Fig. \ref{unc_sensitivity} shows the separate effects of each type of uncertainty on the gains of the transfer function from $\boldsymbol{\dot{{\omega}}}\{1\}$ to $\mathbf{T}_{e x t, G_{1}}\{1\}$ regarding the open loop system for moment {\normalsize \textcircled{\scriptsize 2}}.

A new uncertain closed-loop interconnection $\hat{\mathscr{M}}_{unc}$ is obtained using the controller $\widehat{\mathbf{K}}$ synthesized in the
previous section, which is given by $\hat{\mathscr{M}}_{unc}=\mathcal{F}_{u}\left(\mathscr M, \boldsymbol{\Delta}_{\hat{\mathcal{P}}}\right)$, with $\boldsymbol{\Delta}_{\hat{\mathcal{P}}}=\operatorname{diag}\left(\boldsymbol{\Delta}_{\mathcal{P}}, \boldsymbol{\Delta}_{add}, \boldsymbol{\Delta}_{mul}\right)$. A new array of 330 different plant models $\hat{{\mathscr M}}_{330}$ is obtained from $\hat{\mathscr{M}}_{unc}$. Similarly to how ${\mathscr M}_{200}$ was accomplished, $\hat{{\mathscr M}}_{330}$ results from replacing $\theta_{\bullet}$, $\alpha_{\bullet}$, ${\delta}_{\mathcal{C}_{1}}$ and ${\delta}_{\mathcal{C}_{2}}$ accordingly along 330 equally distributed waypoints in the time domain. In this case, the block $\boldsymbol{\Delta}_{tot}=\operatorname{diag}\left(\boldsymbol{\Delta}_{real}, \boldsymbol{\Delta}_{add}, \boldsymbol{\Delta}_{mul}\right)$ remains uncertain. 

\begin{figure}[!ht]
\centering
 \includegraphics[width=1\textwidth]{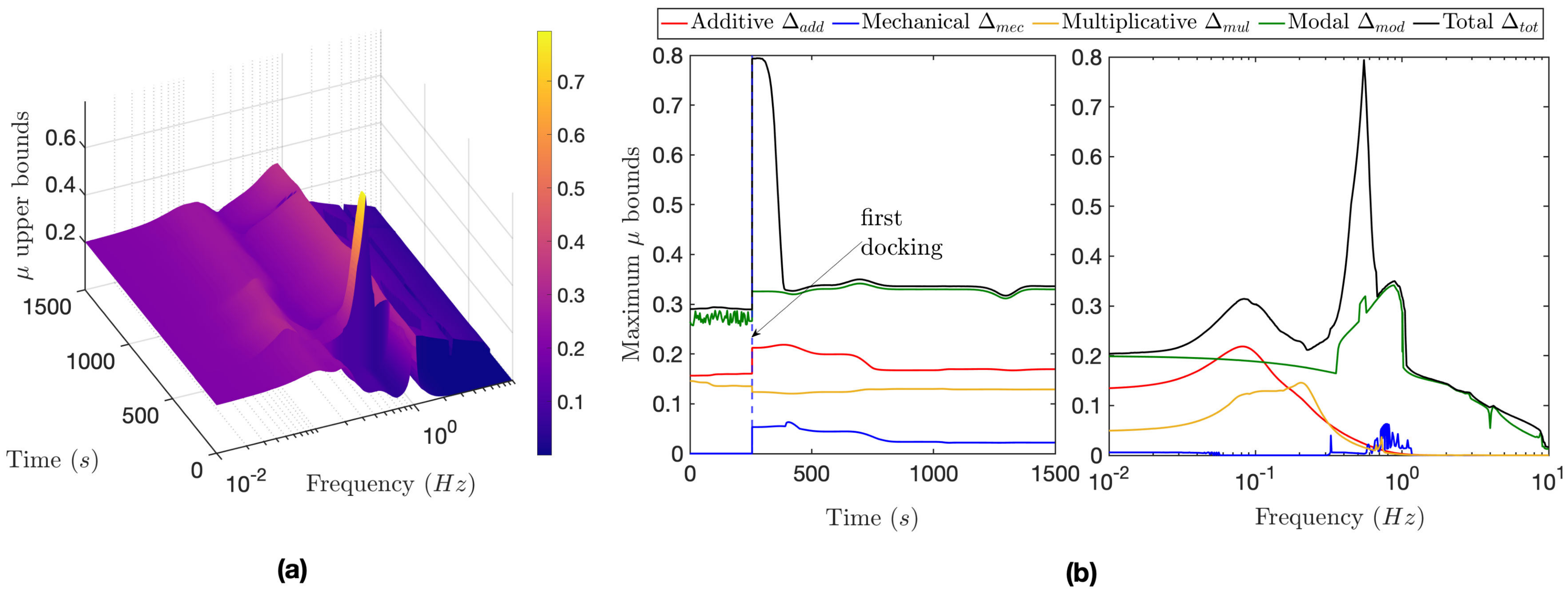}
\caption{Robust stability plots: (a) upper bound across a dense grid of frequencies and different geometrical configurations. (b) side views of the upper bounds computed with respect to different subsets of uncertainty.}
\label{muanalysis} 
\end{figure}

The robust stability of $\hat{{\mathscr M}}_{330}$ is finally evaluated. This assessment is done by calculating the bounds on the structured singular value $\mu_{\delta}$ across a dense grid of frequencies $\omega_{\mu}$, while taking into account all the uncertain 330 plant models. Fig. \ref{muanalysis}a depicts the upper bounds of $\mu_{\delta}$ for the complete set of considered uncertainties $\boldsymbol{\Delta}_{tot}$. Furthermore, Fig. \ref{muanalysis}b illustrates the side views of the upper bounds of the same function with respect to several different sets of uncertainty. The peak of $\mu_{\delta}$, which corresponds to the minimum in stability margin, occurs around $t=275$ $s$, which means it happens right after the first docking. Furthermore, this peak also occurs for a frequency of 0.55 $Hz$, which corresponds to the worst-case $\omega_{1_{\mathcal{S\!A}_{3,4}}}$. However, even when combining all the uncertainty, $\mu_{\delta}$ remains below 0.79 and therefore the loop can tolerate an increase in the uncertainty $\boldsymbol{\Delta}_{tot}$ of 26\% while maintaining stability.

\begin{figure}[!ht]
\centering
 \includegraphics[width=1\textwidth]{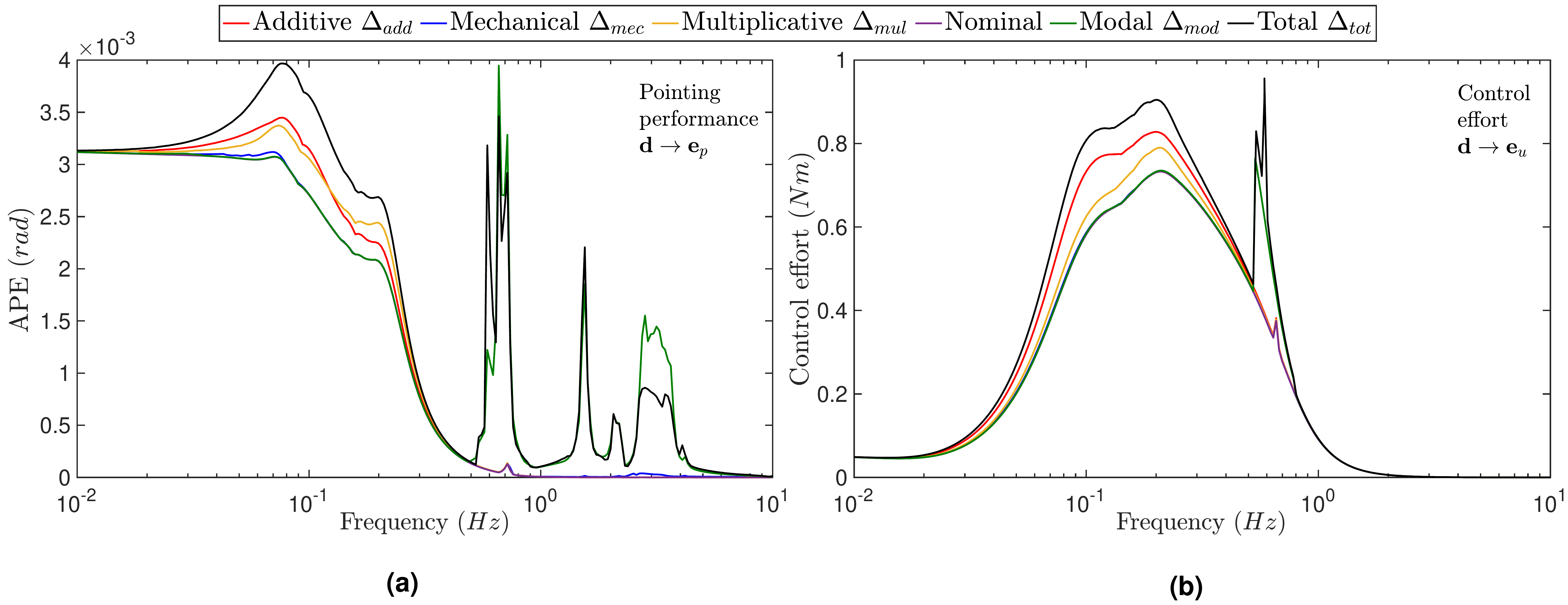}
\caption{Upper bounds on the gains of different performance channels with respect to different uncertainty sets: (a) the absolute pointing error tracking channel $\mathbf{d}\rightarrow \mathbf{e}_{p}$. (b) the control effort channel $\mathbf{d}\rightarrow \mathbf{e}_{u}$.}
\label{performance} 
\end{figure}

The impact of all the subsets of uncertainties was also assessed on different performance indicators using structured singular value computations. Fig. \ref{performance} illustrates the upper bounds on the peak gain for different performance signals across frequencies $\omega_{\mu}$ and for $\hat{{\mathscr M}}_{330}$. The first performance transfer $\mathbf{d}\rightarrow \mathbf{e}_{p}$ can be observed in Fig. \ref{performance}a, which corresponds to the absolute pointing error tracking channel. In this case, the highest peak happens for frequencies around $0.08$ $Hz$. This peak is mainly caused by the worst-case $\boldsymbol{\Delta}_{add}$ and $\boldsymbol{\Delta}_{mul}$, which causes an increase in the gain of the closed-loop system for frequencies around $0.08$ $Hz$. For $\omega_{\mu}>0.5$ $Hz$, the worst-case gains are mainly sensitive to modal uncertainties $\boldsymbol{\Delta}_{mod}$. The peaks that occur at 0.59, 0.66 and 0.72 $Hz$ are once again caused by the uncertain natural frequencies $\omega_{1_{\mathcal{S\!A}_{3,4}}}$. For $\omega_{\mu}>1$ $Hz$, some small peaks can be observed in Fig. \ref{performance}a around the flexible modes' natural frequencies. However, they do not compromise the tracking performance. 

The worst-case gains of a second performance channel $\mathbf{d}\rightarrow \mathbf{e}_{u}$ corresponding to the maximum control effort are shown in Fig. \ref{performance}b. It can be observed that the channel maintains values close to nominal ones even in the presence of significant model uncertainty. However, this channel is also slightly sensitive to the presence of modal uncertainties $\boldsymbol{\Delta}_{mod}$ for frequencies $\omega_{\mu}$ around $\omega_{1_{\mathcal{S\!A}_{3,4}}}$. Indeed, Fig. \ref{performance}b clearly shows that the highest peak happens around $\omega_{1_{\mathcal{S\!A}_{3,4}}}$. Afterwards, there is a visible roll-off, which is caused by $\mathbf{F}_{ro}$. The increase in control effort around $0.08$ $Hz$ is once again caused by the worst-case $\boldsymbol{\Delta}_{add}$ and $\boldsymbol{\Delta}_{mul}$, which forces the controller to work harder around this frequency.

\subsubsection{Stability analysis for the first docking phase}

Fig. \ref{sigma_plot_springmass} has shown that taking into account the dynamic behaviour of the docking mechanisms when modeling such a complex system is of paramount importance. Let us now consider the case of Fig. \ref{sigma_plot_springmass}a, where the robotic arm is docking to the target, since {\Large \textcircled{\normalsize 7}} represents the most critical docking phase. For that reason, the interconnection displayed in Fig. \ref{controldesign}a is now derived by considering the global LFR representation shown in Fig. \ref{compact_lft_springmass}a.

A new uncertain closed-loop interconnection ${\hat{\mathscr{M}}}^7_{unc}$ is obtained using the controller $\widehat{\mathbf{K}}$ synthesized in the
previous section, which is given by ${\hat{\mathscr{M}}}^7_{unc}=\mathcal{F}_{u}\left({\mathscr M}^7, 
\boldsymbol{\Delta}_{\hat{\mathcal{P}}^7}\right)$, with $\mathscr M^7=\mathcal{F}_{l}\left(\mathcal{P}^7_{tot}, \widehat{\mathbf{K}}\right)$ and  $\boldsymbol{\Delta}_{\hat{\mathcal{P}}^7}=\operatorname{diag}\left(\boldsymbol{\Delta}_{\mathcal{P}^7}, \boldsymbol{\Delta}_{add}, \boldsymbol{\Delta}_{mul}\right)$. A new array of 300 different plant models $\hat{\mathscr M}^7_{300}$ is obtained from ${\hat{\mathscr{M}}}^7_{unc}$ by replacing $\delta_{{K}_{1}}^{\bullet}$ with 300 systematically increasing values from 0.1 to $1e^5$ $([N/m]$ or $[Nm/rad])$ and by setting the damping coefficients of $\mathcal{SM}_1$ as $\delta_{{D}_{1}}^{\bullet}=100$ $([Ns/m]$ or $[Nms/rad])$. Furthermore, $\alpha_{\bullet}$ and $\theta_{\bullet}$ are constant for all the 300 different plant models and set according to the system's geometrical configuration displayed in {\Large \textcircled{\normalsize 7}}. 

\begin{figure}[!ht]
\centering
 \includegraphics[width=1\textwidth]{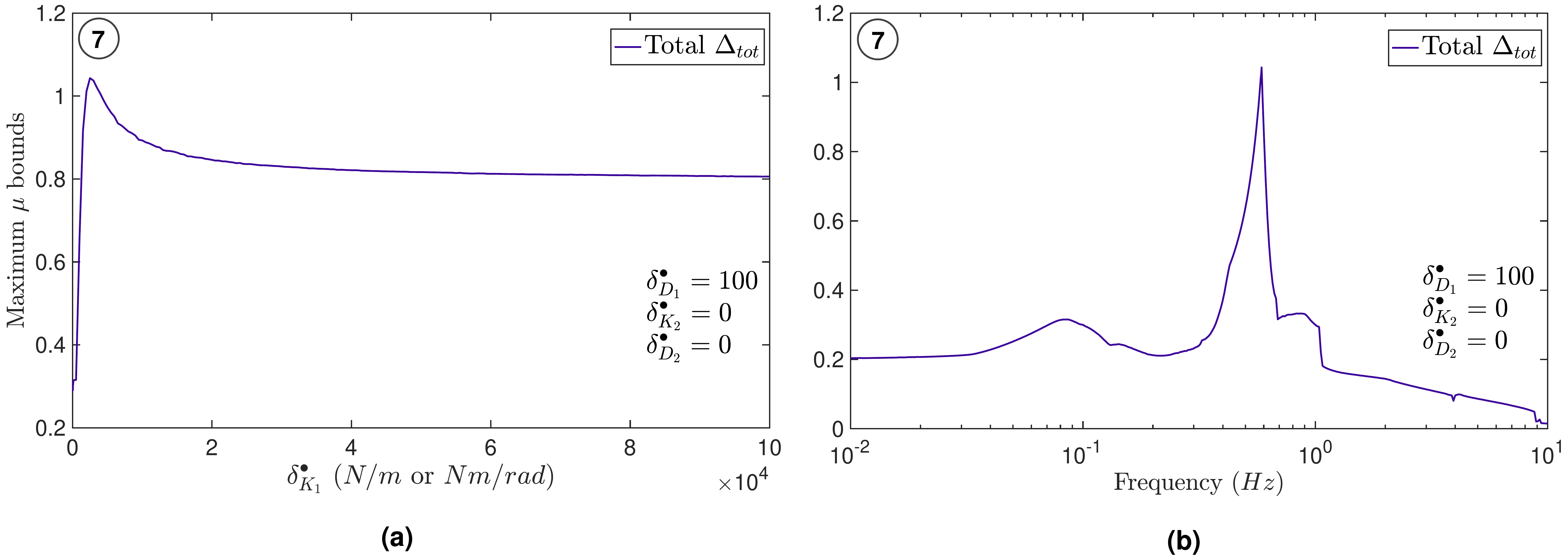}
\caption{Robust stability plots: side views of the upper bounds computed with respect to $\boldsymbol{\Delta}_{tot}$ (Note: The units of $\delta_{{K}_{\bullet}}^{\bullet}$ are $[N/m]$ or $[Nm/rad]$ and the units of $\delta_{{C}_{\bullet}}^{\bullet}$ are $[Ns/m]$ or $[Nms/rad]$, depending on whether the stiffness and damping coefficients are shear or torsional).}
\label{mu_analysis_SM} 
\end{figure}

Fig. \ref{mu_analysis_SM} depicts the side views of the upper bounds of $\mu_{\delta}$ with respect to $\boldsymbol{\Delta}_{tot}$. The peak of $\mu_{\delta}$ occurs around $\delta_{{K}_{1}}^{\bullet}=2.5e^3$ $([N/m]$ or $[Nm/rad])$ as well as for a frequency equal to 0.58 $Hz$, which corresponds once again to the worst-case $\omega_{1_{\mathcal{S\!A}_{3,4}}}$. This peak of $\mu_{\delta}$ is equal to 1.04, what means that the closed-loop can only tolerate 96\% of the considered uncertainty $\boldsymbol{\Delta}_{tot}$ while maintaining stability. This stability analysis thus shows the importance of being able to study the behaviour of docking mechanisms, by having spring-damper systems parameterized according to their stiffness and damping characteristics. These LFR models can be taken into account when designing a controller, so that the closed-loop system does not go unstable when docking takes place. Furthermore, these models can also be used for worst-case analysis and controller validation, as it was demonstrated in this section.

This structured singular value analysis can thus be carried out to perform this type of preliminary Validation and Verification cycles on the linearized model of the system. The obtained results can then be utilized to run a guided Monte Carlo simulation where the most problematic areas are explored in more detail. The sensitivity analysis depicted in Fig. \ref{unc_sensitivity} can also be used to understand which uncertainties have more impact on the dynamic behaviour of the system. In addition, these results can be used to inform the control and system design process in order to perform quick design iterations. It must be noted that the method described in this paper expects the system to undergo small variations around the linearization point, so that the dynamics remain fairly linear. Since most space missions tend to avoid large deflections and nonlinearities in the structural dynamics by design, this approach covers a vast range of applications. In case relevant nonlinearities are present, which can result from the flexible structures undergoing big deflections, this effect has to be considered during the tuning and validation procedures.

\section{Conclusion}

This paper outlined a full modeling and control design methodology for an on-orbit servicing scenario. The presented framework shows how to build a very compact representation of the system by taking into account all the elements which make OOS missions so complex and challenging, namely the coupled flexible spacecraft and robotic arm interactions. This modeling process also introduces a new approach using two spring-damper systems with local uncertain damping and stiffness, offering the possibility of modeling the dynamic behaviour of a docking mechanism and also a closed-loop kinematic chain when the robotic arm configuration is static and perfectly known. The controller synthesis procedure includes a thorough description of how to assemble the design model, including all the details related to the different requirements and limits of performance. The posterior robust performance assessment which was performed is a necessary step to ensure the safety and reliability of the proposed control law. The precious information that can be extracted from this analysis can be used to inform the control and system design process. Afterwards, it can also be utilized to perform quick design iterations. This can lead to structural design adjustments or even to the optimization of certain mechanical parameters, like the mass or the inertial properties of a spacecraft. In addition, the possibility to have these Validation and Verification cycles without the need of high computational burden simulations in such a preliminary phase is extremely important.

\begin{table} [!ht]

	\caption{Chaser and target spacecraft mechanical data. \textbf{Nomenclature}: MoI (Moment of Inertia); PoI (Product of Inertia); CoM (Center of Mass).}
	\label{tab:Sat_prop}	
	\centering
	\resizebox{\textwidth}{!}{
	\begin{tabular}{p{1.5cm} l l r }
		\toprule
		& \textbf{Parameter} & \textbf{Description}  & \textbf{Value and Uncertainty} \\
		\midrule
		\multirow{5}{1.5cm}{\centering Chaser's rigid hub $\mathcal{\mathcal{RH}}_{1}$}
		& $\overrightarrow{G_{1}  P_{1,2}}$ & distance vector between $G_{1}$ and $P_{1,2}$ written in $\mathcal{R}_{\mathcal{RH}_{1}}$ & $[0,\, \pm 0.4365,\, 0]$ m \\
		& $\overrightarrow{G_{1}  J_{0}}$ & distance vector between $G_{1}$ and $J_{0}$ written in $\mathcal{R}_{\mathcal{RH}_{1}}$ & $[0.6508,\, 0,\,  -0.4020]$ m \\
		&  $m^{{\mathcal{RH}}_{1}}$ & mass of $\mathcal{RH}_{1}$ & $188.5\ kg$ \\
		& $\begin{bmatrix}
			\textbf{J}_{xx_{{\mathcal{RH}}_{1}}} & \textbf{J}_{xy_{{\mathcal{RH}}_{1}}} & \textbf{J}_{xz_{{\mathcal{RH}}_{1}}} \\
			 &  \textbf{J}_{yy_{{\mathcal{RH}}_{1}}} & \textbf{J}_{yz_{{\mathcal{RH}}_{1}}} \\
			 &  &  \textbf{J}_{zz_{{\mathcal{RH}}_{1}}} \\
		\end{bmatrix}$ & inertia of $\mathcal{\mathcal{RH}}_{1}$ at $G_{1}$ written in $\mathcal{R}_{\mathcal{RH}_{1}}$ frame & $\begin{bmatrix}
		41.98 & 3.84 & 0 \\
		 &  43.89 & 0 \\
		 &  &  42.64 \\
		\end{bmatrix}\, kg \, m^2 $ \\ \midrule
		
		\multirow{10}{1.5cm}{\centering Chaser's solar arrays ${\mathcal{SA}_{1,2}}$}
		& $\overrightarrow{P_{1,2}  S_{1,2}}$ & distance vector between $P_{1,2}$ and $S_{1,2}$ written in $\mathcal{R}_{\mathcal{SA}_{1,2}}$ & $[0,\, 1.0934,\, 0.0014] \ m$ \\
		& $m^{\mathcal{S\!A}_{1,2}}$ & mass of $\mathcal{SA}_{1,2}$ & $88.93\ kg$ \\
		& $\begin{bmatrix}
		\textbf{J}_{xx_{\mathcal{S\!A}_{1,2}}} & \textbf{J}_{xy_{\mathcal{S\!A}_{1,2}}} & \textbf{J}_{xz_{\mathcal{S\!A}_{1,2}}} \\
		&  \textbf{J}_{yy_{\mathcal{S\!A}_{1,2}}} & \textbf{J}_{yz_{\mathcal{S\!A}_{1,2}}} \\
		&  &  \textbf{J}_{zz_{\mathcal{S\!A}_{1,2}}} \\
		\end{bmatrix}$ & inertia of $\mathcal{SA}_{1,2}$ at $S_{1,2}$ written in $\mathcal{R}_{\mathcal{S\!A}_{1,2}}$ frame 
		& $\begin{bmatrix}
		33.0918 & 0 & 0 \\
		&  7.3819   & -0.0002 \\
		&  &  40.4578 \\
		\end{bmatrix}\, kg\ m^2 $ \\
		& $[\omega_{1_{\mathcal{S\!A}_{1,2}}},\omega_{2_{\mathcal{S\!A}_{1,2}}},\omega_{3_{\mathcal{S\!A}_{1,2}}},\omega_{4_{\mathcal{S\!A}_{1,2}}},\omega_{5_{\mathcal{S\!A}_{1,2}}},\omega_{6_{\mathcal{S\!A}_{1,2}}} ]$ & flexible modes' frequencies & $[1.2850 \pm 20\%,\ 6.5896, \ 7.5231, \ 9.6937, \ 26.1311, \ 28.2408]\ Hz$ \\ 
		& $[\xi_{1_{\mathcal{S\!A}_{1,2}}},\xi_{2_{\mathcal{S\!A}_{1,2}}},\xi_{3_{\mathcal{S\!A}_{1,2}}},\xi_{4_{\mathcal{S\!A}_{1,2}}},\xi_{5_{\mathcal{S\!A}_{1,2}}},\xi_{6_{\mathcal{S\!A}_{1,2}}} ]$ & flexible modes' damping & $0.01$ \\ 
		& $\textbf{L}_{P_{1,2}}^{\mathcal{SA}_{1,2}}$ & modal participation factors & 
		$\begin{bmatrix}
	    -0.0007  & -0.0078  &  7.8872 &  11.7690  &  0.0005  &  0.0010 \\
        -7.9401  &  0   & 0.0007  & -0.0008  &  0.1089 &  12.1014 \\
        -0.3604 &   0  &  0.0006  &  0.0017 &  -2.6631 &   0.5399 \\
        0.0019 &  -0.0066   & 3.9818 &    0.9098 &  -0.0007 &  -0.0033 \\
        0.0272  &  0.0003  & -0.0145 &  -0.0019  &  0.4907  & -0.0221 \\
        -0.0010  &  0.0357  & -2.2185  & -0.2320  & -0.0029 &   0.0012 \\
		\end{bmatrix}$
		\\ \midrule

		\multirow{10}{1.5cm}{\centering Chaser's robotic arm $\mathcal{RA}$}
		& $m^{\mathcal{L}_{\bullet}}$ & mass of $\mathcal{L}_{\bullet}$ & $[4,\, 3.7,\, 8.393,\, 2.275,\, 1.219,\, 1.219,\, 0.1879]\ kg$ \\
		& $x_{\mathcal{L}_{\bullet}}$ & x-coordinate of the CoM of $\mathcal{L}_{\bullet}$ written in $\mathcal{R}_{\mathcal{L}_{\bullet}}$ frame & $[0,\, 0,\, 0.28,\, 0.25,\, 0,\, 0,\, 0]\ m$ \\
		& $y_{\mathcal{L}_{\bullet}}$ & y-coordinate of the CoM of $\mathcal{L}_{\bullet}$ written in $\mathcal{R}_{\mathcal{L}_{\bullet}}$ frame  & $[0,\, 0,\, 0,\, 0,\, 0,\, 0,\, 0]\ m$ \\
		& $z_{\mathcal{L}_{\bullet}}$ & z-coordinate of the CoM of $\mathcal{L}_{\bullet}$ written in $\mathcal{R}_{\mathcal{L}_{\bullet}}$ frame  & $[0,\, 0,\, 0,\, 0,\, 0,\, 0,\, 0]\ m$ \\ 
		& $\textbf{J}_{xx_{{\mathcal{L}}_{\bullet}}}$ & first MoI of $\mathcal{L}_{\bullet}$ at the CoM of $\mathcal{L}_{\bullet}$ written in $\mathcal{R}_{\mathcal{\mathcal{L}}_{\bullet}}$ frame & $[0.0044,\, 0.0103,\, 0.0152,\, 0.0041,\, 0.1112,\, 0.1112,\, 0.0171]\ kg\ m^2$ \\
		& $\textbf{J}_{yy_{{\mathcal{L}}_{\bullet}}}$ & second MoI of $\mathcal{L}_{\bullet}$ at the CoM of $\mathcal{L}_{\bullet}$ written in $\mathcal{R}_{\mathcal{\mathcal{L}}_{\bullet}}$ frame & $[0.0044,\, 0.0103,\, 0.2269,\, 0.0494,\, 0.1112,\, 0.1112,\, 0.0171]\ kg\ m^2$ \\		
		& $\textbf{J}_{zz_{{\mathcal{L}}_{\bullet}}}$ & third MoI of $\mathcal{L}_{\bullet}$ at the CoM of $\mathcal{L}_{\bullet}$ written in $\mathcal{R}_{\mathcal{\mathcal{L}}_{\bullet}}$ frame & $[0.0072,\, 0.0067,\, 0.2269,\, 0.0494,\, 0.2194,\, 0.2194,\, 0.0338]\ kg\ m^2$ \\
		& $\textbf{J}_{xy_{{\mathcal{L}}_{\bullet}}}$ & first PoI of $\mathcal{L}_{\bullet}$ at the CoM of $\mathcal{L}_{\bullet}$ written in $\mathcal{R}_{\mathcal{\mathcal{L}}_{\bullet}}$ frame & $[0,\, 0,\, 0,\, 0,\, 0,\, 0,\, 0]\ kg\ m^2$ \\
		& $\textbf{J}_{xz_{{\mathcal{L}}_{\bullet}}}$ & second PoI of $\mathcal{L}_{\bullet}$ at the CoM of $\mathcal{L}_{\bullet}$ written in $\mathcal{R}_{\mathcal{\mathcal{L}}_{\bullet}}$ frame & $[0,\, 0,\, 0,\, 0,\, 0,\, 0,\, 0]\ kg\ m^2$ \\		
		& $\textbf{J}_{yz_{{\mathcal{L}}_{\bullet}}}$ & third PoI of $\mathcal{L}_{\bullet}$ at the CoM of $\mathcal{L}_{\bullet}$ written in $\mathcal{R}_{\mathcal{\mathcal{L}}_{\bullet}}$ frame & $[0,\, 0,\, 0,\, 0,\, 0,\, 0,\, 0]\ kg\ m^2$ \\	\midrule
		
		\multirow{5}{1.5cm}{\centering Target's rigid hub $\mathcal{\mathcal{RH}}_{2}$}
		& $\overrightarrow{G_{2}  P_{3,4}}$ & distance vector between $G_{2}$ and $P_{3,4}$ written in $\mathcal{R}_{\mathcal{RH}_{2}}$ & $[0,\, \pm 0.3395,\, 0]$ m \\
		&  $m^{{\mathcal{RH}}_{2}}$ & mass of $\mathcal{RH}_{2}$ & $24.96 \pm 10\%\ kg$ \\
		& $\begin{bmatrix}
			\textbf{J}_{xx_{{\mathcal{RH}}_{2}}} & \textbf{J}_{xy_{{\mathcal{RH}}_{2}}} & \textbf{J}_{xz_{{\mathcal{RH}}_{2}}} \\
			 &  \textbf{J}_{yy_{{\mathcal{RH}}_{2}}} & \textbf{J}_{yz_{{\mathcal{RH}}_{2}}} \\
			 &  &  \textbf{J}_{zz_{{\mathcal{RH}}_{2}}} \\
		\end{bmatrix}$ & inertia of $\mathcal{RH}_{2}$ at $G_{2}$ written in $\mathcal{R}_{\mathcal{RH}_{2}}$ frame & $\begin{bmatrix}
		 2.684 \pm 10\% & 0.058 & 0.054 \\
		 &  2.012 \pm 10\% & -0.104 \\
		 &  &  2.32 \pm 10\% \\
		\end{bmatrix}\, kg \, m^2 $ \\ \midrule
		
		\multirow{10}{1.5cm}{\centering Target's solar arrays ${\mathcal{SA}_{3,4}}$}
		& $\overrightarrow{P_{3,4}  S_{3,4}}$ & distance vector between $P_{3,4}$ and $S_{3,4}$ written in $\mathcal{R}_{\mathcal{SA}_{3,4}}$ frame & $[0,\, 0.7446,\, 0] \ m$ \\
		& $m^{\mathcal{S\!A}_{3,4}}$ & mass of ${\mathcal{SA}_{3,4}}$ & $11.3497\ kg$ \\
		& $\begin{bmatrix}
		\textbf{J}_{xx_{\mathcal{S\!A}_{3,4}}} & \textbf{J}_{xy_{\mathcal{S\!A}_{3,4}}} & \textbf{J}_{xz_{\mathcal{S\!A}_{3,4}}} \\
		&  \textbf{J}_{yy_{\mathcal{S\!A}_{3,4}}} & \textbf{J}_{yz_{\mathcal{S\!A}_{3,4}}} \\
		&  &  \textbf{J}_{zz_{\mathcal{S\!A}_{3,4}}} \\
		\end{bmatrix}$ & inertia of ${\mathcal{SA}_{3,4}}$ at $S_{3,4}$ written in $\mathcal{R}_{\mathcal{SA}_{3,4}}$ frame
		& $\begin{bmatrix}
		1.9566 & 0 & 0 \\
		&  0.3404   & 0 \\
		&  &  2.2968 \\
		\end{bmatrix}\, kg\ m^2 $ \\
		& $[\omega_{1_{\mathcal{S\!A}_{3,4}}},\omega_{2_{\mathcal{S\!A}_{3,4}}},\omega_{3_{\mathcal{S\!A}_{3,4}}},\omega_{4_{\mathcal{S\!A}_{3,4}}},\omega_{5_{\mathcal{S\!A}_{3,4}}},\omega_{6_{\mathcal{S\!A}_{3,4}}} ]$ & flexible modes' frequencies & $[0.6493 \pm 20\%,\  2.2480, \ 3.9870, \ 4.3455, \  10.9601, \ 18.2744]\ Hz$ \\ 
		& $[\xi_{1_{\mathcal{S\!A}_{3,4}}},\xi_{2_{\mathcal{S\!A}_{3,4}}},\xi_{3_{\mathcal{S\!A}_{3,4}}},\xi_{4_{\mathcal{S\!A}_{3,4}}},\xi_{5_{\mathcal{S\!A}_{3,4}}},\xi_{6_{\mathcal{S\!A}_{3,4}}} ]$ & flexible modes' damping & $0.01$ \\

		& $\textbf{L}_{P_{3,4}}^{\mathcal{SA}_{3,4}}$ & modal participation factors & 
		$\begin{bmatrix}
	    0.0003 &  0 &   -2.7332 &  -2.8462 &   0.0001 &  -0.0003 \\
        2.8655 &   0   & 0  & 0.0002  & -0.0025  & -2.9305 \\
        -0.0002 &  0  & -1.5206  & -0.3709  &  0.0022  &  0.0003 \\
        -0.0119   & 0 &  -0.0058 &  -0.0017  & -0.5800  &  0.0123 \\
        0  & 0  &  0.8207 &   0.0958  &  0.0002  & -0.0001 \\
        0.0008   & 0.0001  & -0.0007  &  0  &  0.0596  & -0.0009 \\
		\end{bmatrix}$
		\\ \midrule

	\end{tabular}
	
}
\end{table}

\section*{Funding Sources}

This work was funded by ISAE-SUPAERO.




\bibliographystyle{model1-num-names}
\bibliography{library.bib}







\end{document}